# Three-Dimensional Multiphysics Modelling of Helicon Wave Heating and Antenna–Plasma Coupling for Boundary Density Control in Toroidal Fusion Plasmas

Hua Zhou[1,2], Lei Chang[3*], Guo-Sheng Xu[1,2*], Yi-Wei Zhang[1,2], Matthew Hole[4], Dan Du[5], Zhi-Song Qu[6], Mu-Quan Wu[7]

[1]Institute of Plasma Physics, HFIPS, Chinese Academy of Sciences, Hefei 230031, China

[2]Science Island Branch of Graduate, University of Science and Technology of China, Hefei 230026, China

[3]Fundamental Plasma Physics and Innovative Applications Laboratory, School of Electrical Engineering, Chongqing University, Chongqing 400044, China

[4]Mathematical Sciences Institute, Australian National University, Canberra, ACT 0200, Australia

[5]Department of Mathematics and Physics, University of South China, Hengyang 421001, China

[6]School of Physical and Mathematical Sciences, Nanyang Technological University, Singapore 637371, Singapore

[7]College of Physics and Optoelectronic Engineering, Shenzhen University, Shenzhen 518060, China

[*]Email: leichang@cqu.edu.cn (Lei Chang), gsxu@ipp.ac.cn (Guo-Sheng Xu)

**Abstract:** Active control of scrape-off-layer (SOL) density is emerging as a critical requirement for improving ion cyclotron resonance heating (ICRH) and enabling high-performance steady-state operation in future magnetic-confinement fusion devices. Helicon wave excitation offers a promising physics-based approach to generating high-density boundary plasmas with high ionization efficiency and low impurity release. In this work, we develop THEMIS (Toroidal Helicon ElectroMagnetic Integrated Solver) model and code, a fully three-dimensional (3D) multiphysics model of helicon wave propagation and power deposition in a toroidal fusion-relevant configuration, employing a finite-temperature thermal dielectric tensor and full-vessel electromagnetic simulation constrained by experimentally measured magnetic-field and density profiles. The code quantifies the relative contributions of Doppler-shifted cyclotron damping, anomalous Doppler damping, collisional damping, and Landau damping, and demonstrates that slow-wave (Trivelpiece–Gould/TG) propagation and electron Landau damping dominate the accessible heating regime in Helimak device. A comparative study of four planar antenna geometries under the present protruding-window configuration shows that geometric cutoffs and SOL density gradients severely limit power penetration into the core-accessible region. To address this constraint, we introduce a recessed-window launch scheme that positions the dielectric window inside the vacuum vessel and perform systematic parameter scans of window position, antenna geometry, and installation orientation. From these analyses, we identify the key physics-driven principles governing efficient helicon wave coupling in toroidal geometry: the importance of open-circuit termination, maximized strap length and width for enhanced slow-wave excitation, controlled inter-turn spacing, and sufficient clearance from metallic walls to avoid near-field suppression. Guided by these principles, we designed an optimized racetrack spiral antenna that increases coupling efficiency by more than an order of magnitude compared with conventional short-circuited rectangular spiral antenna. The results provide new insights into helicon wave interaction with toroidal boundary plasmas and establish a validated antenna–window co-design strategy for future helicon-assisted SOL density control and RF coupling enhancement in tokamaks.

## 1. Introduction

The ability to regulate edge and scrape-off-layer (SOL) plasma conditions is central to achieving high-performance and steady-state operation in magnetic-confinement fusion devices. The SOL plays a decisive role in determining radio-frequency (RF) wave accessibility and absorption, impurity transport, and divertor power handling, and thus directly impacts global confinement and reactor sustainment strategies [1-3]. Ion cyclotron resonance heating (ICRH) remains one of the most versatile RF heating systems in tokamaks and stellarators, offering high-power density, established technological maturity, and direct ion heating under reactor-relevant conditions [4-6]. However, effective ICRH operation is frequently limited by the presence of low-density boundary plasmas, in which the fast wave becomes evanescent, leading to enhanced reflection, impaired coupling efficiency, and non-uniform antenna loading [7-9]. These challenges are expected to persist—and potentially worsen—in next-generation devices with large magnetic field and plasma size, such as ITER and DEMO. To mitigate fast-wave evanescence, present machines commonly employ local gas puffing near ICRH antennas to transiently increase the edge density [10-13]. While this approach can improve near-field matching, it introduces several drawbacks including enhanced radiative losses, difficulties in achieving spatially uniform density profiles, and limited temporal controllability due to neutral-gas transport dynamics. These limitations motivate the search for alternative, active boundary-density control methods that do not rely on additional neutral sources and that can operate in a steady and localized manner.

Helicon wave excitation constitutes a promising candidate for such control. Helicon waves, lying between the ion and electron cyclotron frequencies, can drive exceptionally efficient ionization and produce high-density plasmas in an electrode-less configuration [14-17]. They have been extensively employed in plasma-material interaction studies [18-21] and space electric propulsion [22-26], and recent investigations suggest their ability of heating and current drive in fusion experiments [27-32]. The combination of high ionization efficiency, low impurity production, and strong slow-wave (Trivelpiece–Gould or TG) components indicates that helicon waves could be exploited to raise the SOL density and thereby enhance ICRH coupling in toroidal fusion devices. If validated under fusion-relevant conditions, helicon-assisted SOL control may further contribute to reducing antenna loading asymmetries, alleviating divertor heat fluxes, and influencing edge stability, including edge-localized modes (ELMs).

To examine this concept, we develop THEMIS (Toroidal Helicon ElectroMagnetic Integrated Solver), inspired by the Hybrid Circuit/3DLHDAP code [33]. This model is a three-dimensional(3D) full-wave electromagnetic multiphysics model of helicon wave propagation and power deposition referring to the Helimak, a toroidal fusion device [34-35]. The model employs a finite-temperature thermal dielectric tensor, incorporates realistic magnetic-field and density profiles derived from experiment, and resolves multiple damping channels relevant to helicon wave heating. We compare four planar antenna geometries used in initial experiments under the current launch configuration, quantify the physical mechanisms limiting coupling, and identify the dominant slow-wave absorption processes. Guided by this analysis, we introduce a recessed-window design enabling direct wave access to the main plasma and perform systematic parametric scans of antenna geometry, placement, and orientation. From these results, we derive physics-based design principles for next-generation helicon wave antennas aimed at boundary-plasma control in tokamak environments. The outcomes of this study provide theoretical and computational foundation for future helicon-assisted SOL density modulation experiments and their integration with ICRH systems.

## 2. Modelling Framework

### *2.1 Full-Vessel 3D Geometry and Antenna Configuration*

The Helimak device was designed by Professor Stan Luckhardt of the University of California, San Diego (UCSD), with its detailed engineering design carried out by the Institute of Plasma Physics, Chinese Academy of Sciences (ASIPP). The device was assembled and commissioned at the University of Texas at Austin in 2003. In March 2018, the Fusion Research Centre (FRC) at the University of Texas at Austin agreed to donate the Helimak device and its auxiliary systems to Shenzhen University free of charge. Currently, the device has been reassembled and upgraded at Shenzhen University, and initial plasma discharge experiments have been initiated. As the plasma parameters produced in Helimak resemble the physical conditions SOL region of contemporary tokamak devices, it is regarded as an important experimental platform for studying the physics of the SOL region in magnetically confined fusion devices.

We construct a full 3D model of the Helimak vacuum vessel and its helicon wave antenna structures. The existing protruding window forms a waveguide-like cavity. Four antenna geometries—rectangular spiral antenna (RSA), spiral antenna (SA), comb antenna, and S-bend antenna—are incorporated using experimentally measured dimensions. A ceramic window provides the RF–plasma interface. Figure 1 shows a schematic diagram of the 3D full vacuum chamber and the helicon wave antenna in the Helimak device. The main vacuum chamber features a toroidally axisymmetric structure aligned along the z-axis, with a height of 2 m, an inner wall radius of 0.6 m, and an outer wall radius of 1.6 m. The chamber is constructed of stainless steel. The current configuration employs a conventional protruding window. This racetrack-shaped window (0.3 m wide and 0.45 m high) extends outward from the outer wall of the vacuum chamber, forming a waveguide channel with a radial depth of 0.1 m. The helicon wave antenna is mounted externally to the window. A ceramic window (0.19 m high × 0.22 m wide × 0.01 m thick) serves as the interface, through which the helicon wave antenna couples with the main plasma in the Helimak device. The centre of the window is located 0.425 m above the bottom of the device.

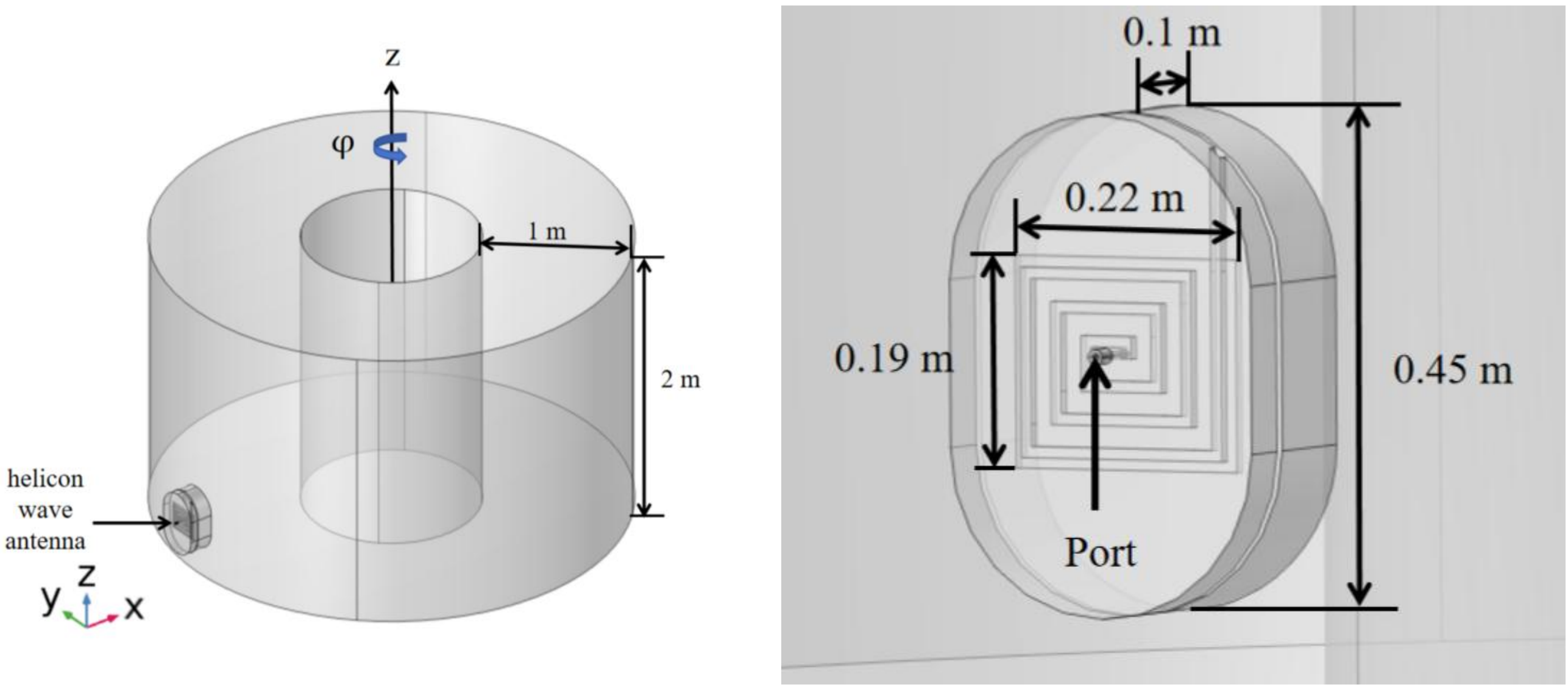

(a) (b)

Figure 1. Schematic illustration of the Helimak configuration used in this study: (a) 3D full vacuum chamber; (b) the helicon wave antenna and waveguide channel.

In 2025, our team conducted the first round of helicon wave heating experiments on the Helimak device, aiming to evaluate the performance of four different antenna types in exciting helicon waves within this toroidal facility. Figure 2 shows schematics of the four antennas used in the study. The antenna is driven by RF power supply of 13.56 MHz. The RF signal is delivered to antenna via a coaxial cable with characteristic impedance of 50 Ω. The antenna termination is grounded to the metallic wall of the Helimak device through a copper strap.

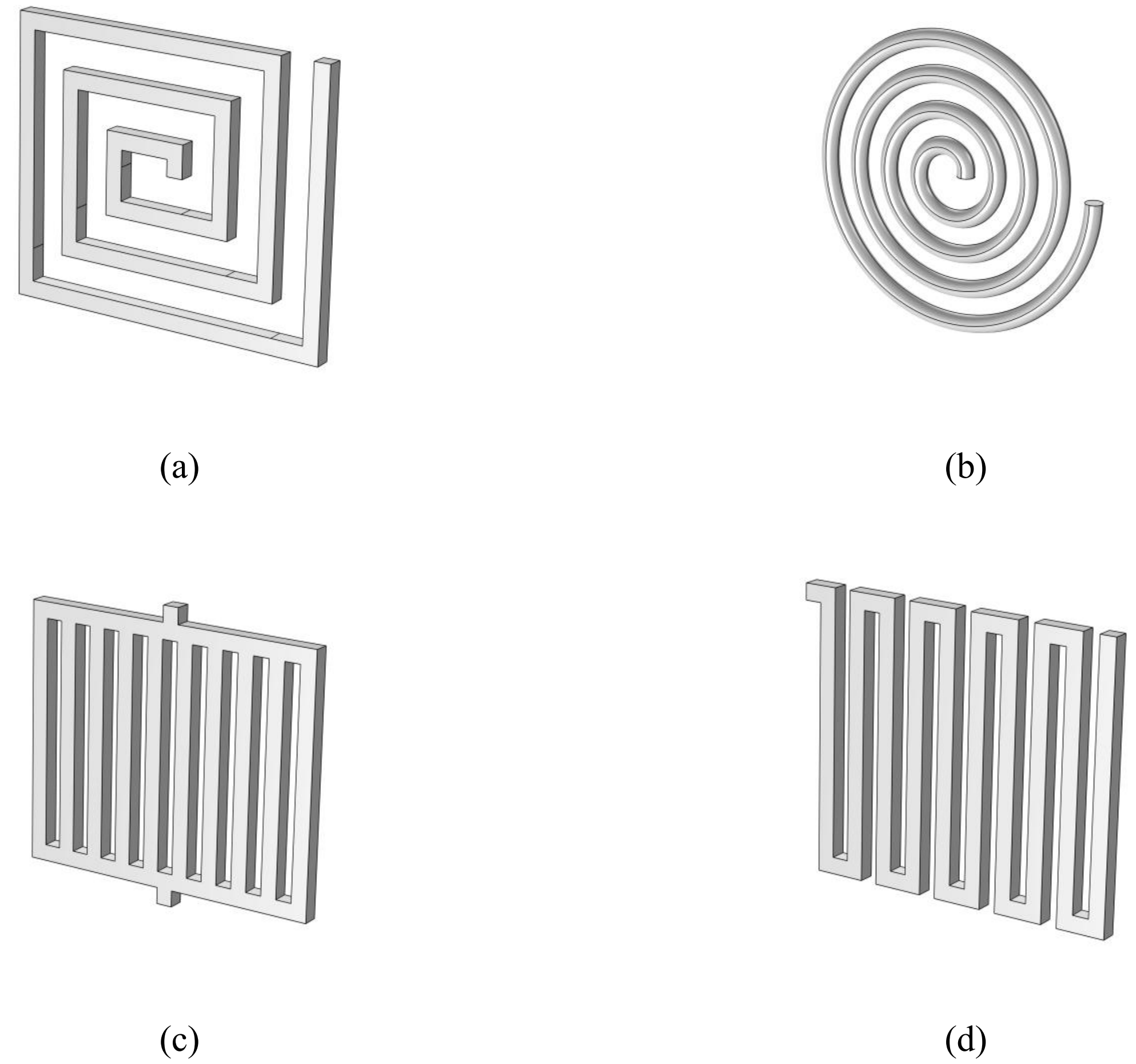

Figure 2. Geometrical schematics of the four antenna configurations investigated in this work: (a) rectangular spiral antenna (RSA), (b) spiral antenna (SA), (c) comb antenna, and (d) S-bend antenna, illustrating their distinct current paths and axial phasing characteristics.

*2.2 Thermal Dielectric Tensor and Wave Equation*

Electromagnetic fields in the magnetized plasma are solved using the thermal dielectric tensor, which includes finite-temperature effects, collisions, and cyclotron harmonics. The model supports both slow-wave (TG) and fast-wave (helicon) branches. Power deposition is decomposed into Doppler-shifted cyclotron, anomalous Doppler, collisional, and Landau damping terms. The wave equation satisfied by electromagnetic waves in magnetized plasma can be derived from Maxwell's equations as follows:

$$\nabla(\nabla\cdot\boldsymbol{E})-\nabla^2\boldsymbol{E}=\omega^2\mu_0\varepsilon_0\overline{\overline{\varepsilon}}_{\mathrm{r}}\cdot\boldsymbol{E} \tag{1}$$

In the formula, $\mu_0$ is vacuum permeability, $\varepsilon_0$ is vacuum permittivity. When the direction of the external magnetic field is along the z-axis, the equivalent relative dielectric tensor $\overline{\overline{\varepsilon}}_{\mathrm{r}}$ of the thermal plasma [36,37] is expressed as:

$$\overline{\overline{\varepsilon}}_{\mathrm{r}}=\begin{bmatrix}\varepsilon_1 & i\varepsilon_2 & 0\\ -i\varepsilon_2 & \varepsilon_1 & 0\\ 0 & 0 & \varepsilon_3\end{bmatrix} \tag{2}$$

$$\varepsilon_1=1+\sum_{\alpha}\frac{\omega_{\mathrm{p\alpha}}^2}{2\omega|k_{\|}|\upsilon_{\mathrm{T\alpha}}}\left[Z(\zeta_{-1}^{\alpha})+Z(\zeta_{1}^{\alpha})\right] \tag{3}$$

$$\varepsilon_2=\sum_{\alpha}\frac{\omega_{\mathrm{p\alpha}}^2}{2\omega|k_{\|}|\upsilon_{\mathrm{T\alpha}}}\left[Z(\zeta_{-1}^{\alpha})-Z(\zeta_{1}^{\alpha})\right] \tag{4}$$

$$\varepsilon_3=1-\sum_{\alpha}\frac{\omega_{\mathrm{p\alpha}}^2Z'(\zeta_0^{\alpha})}{k_{\|}^2\upsilon_{\mathrm{T\alpha}}^2+jv_{\alpha}|k_{\|}|\upsilon_{\mathrm{T\alpha}}Z(\zeta_0^{\alpha})} \tag{5}$$

Here, the subscript α denotes different particles, such as electrons and He ions, the symbol $\upsilon_{\mathrm{T\alpha}}=\sqrt{T_\alpha/m_\alpha}$ is the α particle thermal velocity, $\omega_{\mathrm{p\alpha}}=\sqrt{nq_\alpha^2/(\varepsilon_0 m_\alpha)}$ is the α particle plasma frequency, $\omega_{\mathrm{c\alpha}}=|B|q_\alpha/m_\alpha$ is the α particle cyclotron frequency, $k_{\|}$ is the wave vector component of the parallel magnetic field. The symbol $Z(\zeta_n^\alpha)$ is the dispersion function, and $Z'(\zeta_n^\alpha)$ is the first derivative of the dispersion function. Further, the symbol $\zeta_n^\alpha=(\omega+iv_\alpha+n\omega_{\mathrm{c\alpha}})/(|k_{\|}|\upsilon_{\mathrm{T\alpha}})$ is the argument of the dispersion function, $\omega$ is the frequency of the radio frequency source. Moreover, $v_\alpha$ is the α particle collision frequency, $v_{\mathrm{en}}$, $v_{\mathrm{ei}}$ and $v_{\mathrm{ie}}$ are the electron-neutral particle collision frequency, electron-ion collision frequency, and ion-electron collision frequency respectively. They satisfy the following relationships [16, 38, 39]:

$$v_{\mathrm{en}}\approx1.3\times10^6 p_{\mathrm{n}}\mathrm{T}_e \tag{6}$$

$$v_{\mathrm{ei}}\approx3\times10^{-6}\times\frac{n}{T_{\mathrm{e}}^{3/2}}\times\left[23+\mathrm{In}\left(\frac{T_{\mathrm{e}}^{3/2}}{\sqrt{n}}\right)\right] \tag{7}$$

$$v_{ie}\approx5\times10^{-8}\times\frac{1}{\sqrt{m_{\mathrm{i}}}}\times\frac{n}{T_{\mathrm{i}}^{3/2}}\times\left[23+\mathrm{In}\left(\frac{T_{\mathrm{e}}^{3/2}}{\sqrt{n}}\right)\right] \tag{8}$$

In plasmas, helicon waves exhibit multiple energy deposition mechanisms, e.g. Doppler-shifted cyclotron damping (DSCD), anomalous Doppler damping (ADD), collisional damping (CD), and Landau damping (LD). The local power deposition satisfies the following [40]:

$$\begin{aligned} P_{\text{abs}} &= \frac{1}{2}\omega\varepsilon_0 \cdot \operatorname{Im}\left[ \boldsymbol{E}^* \cdot \overline{\overline{\varepsilon}}_{\text{r}} \cdot \boldsymbol{E} \right] \\ &\approx \frac{1}{2}\omega\varepsilon_0 \cdot \operatorname{Im}\left\{ 1+\sum_{\alpha} \frac{\omega_{\text{p}\alpha}^2}{\omega|k_{\text{z}}|\upsilon_{\text{T}\alpha}} Z\left(\zeta_{-1}^{\alpha}\right) \right\} \cdot |E_-|^2 \\ &+ \frac{1}{2}\omega\varepsilon_0 \cdot \operatorname{Im}\left\{ 1+\sum_{\alpha} \frac{\omega_{\text{p}\alpha}^2}{\omega|k_{\text{z}}|\upsilon_{\text{T}\alpha}} Z\left(\zeta_{1}^{\alpha}\right) \right\} \cdot |E_+|^2 \\ &+ \frac{1}{2}\omega\varepsilon_0 \cdot \operatorname{Im}\left\{ 1-\sum_{\alpha} \frac{\omega_{\text{p}\alpha}^2}{\omega(\omega+j\nu_{\alpha})} \right\} \cdot |E_{\parallel}|^2 \\ &+ \frac{1}{2}\omega\varepsilon_0 \cdot \operatorname{Im}\left\{ 1-\sum_{\alpha} \frac{\omega_{\text{p}\alpha}^2}{k_{\text{z}}^2\upsilon_{\text{T}\alpha}^2} \cdot Z'\left(\zeta_{0}^{\alpha}\right) \right\} \cdot |E_{\parallel}|^2 \end{aligned} \tag{9}$$

Where the symbol $E_-=(E_x-iE_y)/\sqrt{2}$ is the right-hand polarized electric field component, and $E_+=(E_x+iE_y)/\sqrt{2}$ is the left-hand polarized electric field component. The first term on the right-hand side of Equation (9) is the power deposition caused by Doppler-shifted cyclotron damping $P_{\text{abs_DSCD}}$, the second term is the power deposition caused by anomalous Doppler damping $P_{\text{abs_ADD}}$, the third term is the power deposition caused by collisional damping $P_{\text{abs_CD}}$, and the fourth term is the power deposition caused by Landau damping $P_{\text{abs_LD}}$. In THEMIS, the forward power ($P_{\text{fwd}}$) of the fixed power source is 1 kW. The actual input power at the antenna port ($P_{\text{port}}=P_{\text{fwd}}$-$P_{\text{ref}}$) and the power absorbed ($P_{\text{abs}}$) by the plasma are calculated using finite element method via the RF module of COMSOL Multiphysics Framework. The absorption efficiency of the plasma is defined as:

$$\eta_{\text{abs}} = \frac{P_{\text{abs}}}{P_{\text{fwd}}} \tag{10}$$

This study does not account for thermal losses from the antenna and metal walls in the model. Based on energy conservation, the input power at the antenna port equals the power absorbed by the plasma. The absorption efficiency is defined as the ratio of the power absorbed by the plasma to the forward wave power in the transmission line. This parameter essentially reflects the reflection characteristics of the antenna: a lower value indicates a higher reflection coefficient and greater difficulty in achieving impedance matching.

### *2.3 Experimental Plasma Profiles and Magnetic Geometry*

The Helimak toroidal magnetic field and radial density profiles are imported from calibrated experimental data. The model includes only radial variations, consistent with the device symmetry. A tensor rotation is applied to align the dielectric tensor with the toroidal magnetic geometry. The toroidal field magnet system of the Helimak consists of sixteen rectangular coils (0.153 m × 0.26 m) arranged along the toroidal direction, with the inhomogeneity of the toroidal magnetic field being less than 1.1%. Currently, Helimak device only uses electron cyclotron resonance heating system to conduct helium gas discharge experiments. The radial distribution of electron density is measured and determined by an array of Langmuir probes and is shown in Figure 3. The magnetic field distribution is provided by a magnetic field program, where the radial distribution of the magnetic field in the plane is presented in Figure 4. For the helicon wave heating study on the Helimak, only the radial variations of electron density and magnetic field are considered, and the magnetic field is assumed to have only the toroidal component. It is assumed that the helium ion density is equal to the electron density, and the helium ion temperature and electron temperature are constant, with values of 10 eV and 1 eV, respectively.

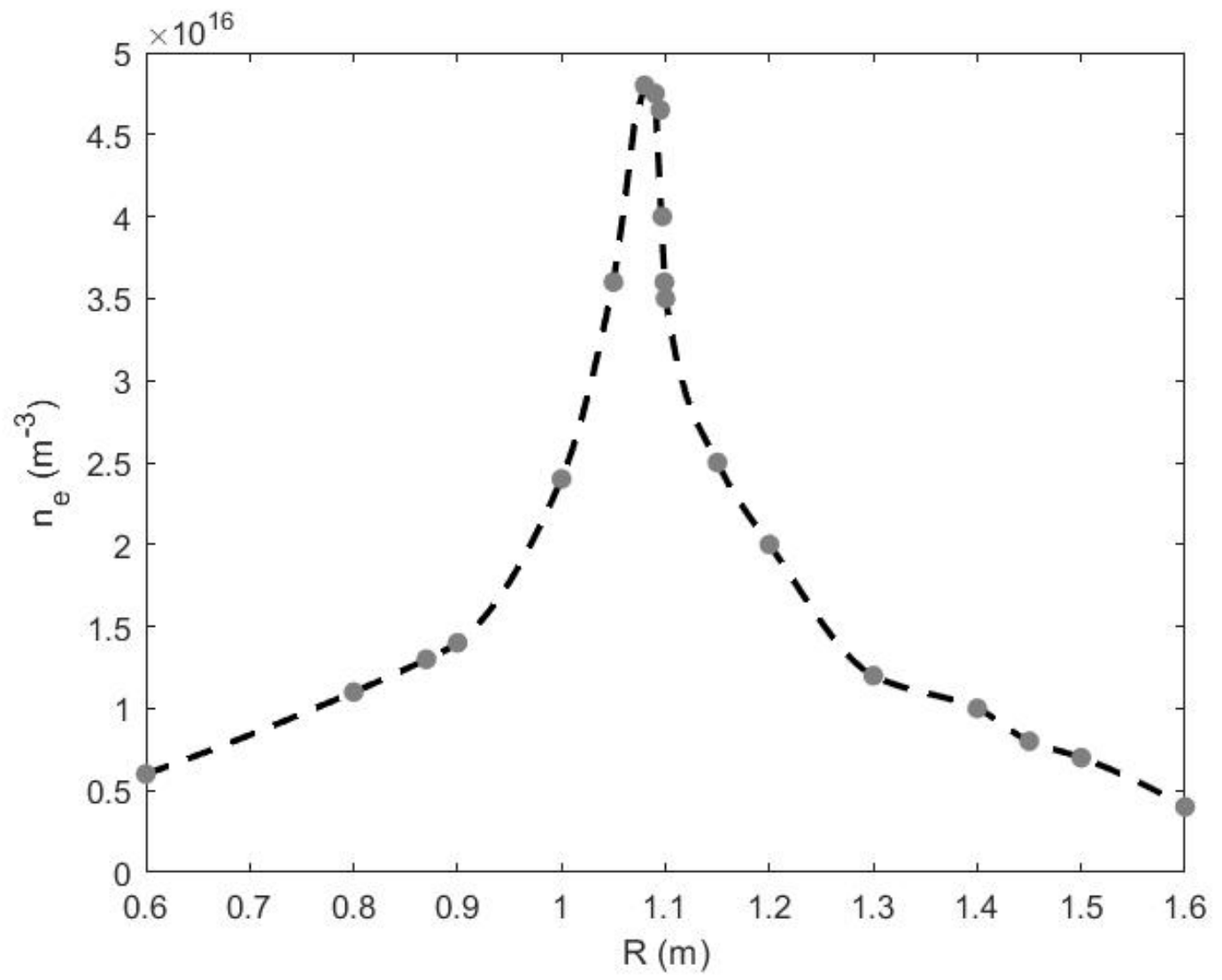

Figure 3. Radial variation of the electron density in the Helimak configuration, serving as the equilibrium density profile for wave propagation and analyses of power deposition.

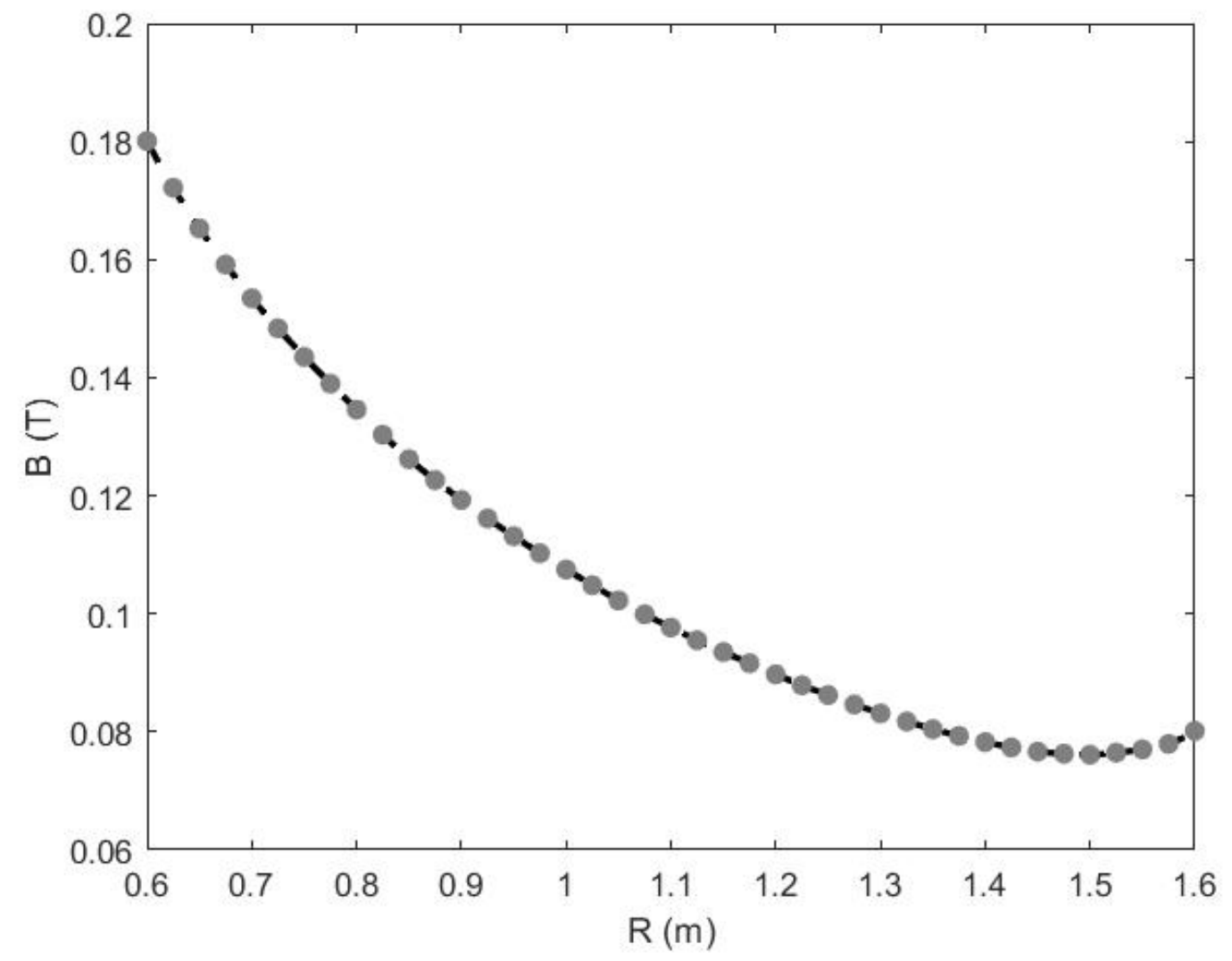

Figure 4. Radial variation of the background toroidal magnetic field in the Helimak configuration, providing the equilibrium magnetic field profile used in the wave and power deposition calculations.

In THEMIS, the thermal dielectric tensor $\overline{\overline{\varepsilon}}'_{\mathrm{r}}$ for the toroidal magnetic field direction is obtained by performing rotational transformation on original magnetized thermal dielectric tensor $\overline{\overline{\varepsilon}}_{\mathrm{r}}$ [41]:

$$\overline{\overline{\varepsilon}}'_{\mathrm{r}} = \mathbf{T} \cdot \overline{\overline{\varepsilon}}_{\mathrm{r}} \cdot \mathbf{T}^{-1} = \begin{pmatrix} \varepsilon'_{xx} & \varepsilon'_{xy} & \varepsilon'_{xz} \\ \varepsilon'_{yx} & \varepsilon'_{yy} & \varepsilon'_{yz} \\ \varepsilon'_{zx} & \varepsilon'_{zy} & \varepsilon'_{zz} \end{pmatrix} \tag{11}$$

The rotation matrix is:

$$\mathbf{T} = \begin{pmatrix} \cos\varphi & 0 & \sin\varphi \\ \sin\varphi & 0 & \cos\varphi \\ 0 & -1 & 0 \end{pmatrix} \tag{12}$$

Where the symbol $\varphi$ is toroidal angle in the Helimak cylindrical coordinate system.

*2.4 Dispersion Relation and Wave Propagation Characteristics*

Using the local plasma parameters, dispersion diagrams are computed for the parallel wavenumber spectra derived from Fourier analysis of each antenna. Slow waves propagate only in low-density regions near the boundary. Fast waves propagate into the core only within a narrow range of small parallel wavenumbers. Assuming the electromagnetic wave is a plane wave, the symbol $\nabla$ in Equation (1) can be replaced by $\vec{k}$ (wave vector), and Equation (1) can be rewritten as:

$$(\vec{k}\vec{k} - k^2 \overline{\overline{I}} + \frac{\omega^2}{c^2} \overline{\overline{\varepsilon}}) \cdot \vec{E} = 0 \tag{13}$$

To ensure the electric field is non-zero, the following condition must hold [36]:

$$\det \begin{vmatrix} \frac{\omega^2}{c^2} S - k_{\parallel}^2 & iD\frac{\omega^2}{c^2} & k_{\perp}k_{\parallel} \\ -iD\frac{\omega^2}{c^2} & \frac{\omega^2}{c^2} S - k_{\perp}^2 - k_{\parallel}^2 & 0 \\ k_{\perp}k_{\parallel} & 0 & \frac{\omega^2}{c^2} P - k_{\perp}^2 \end{vmatrix} = 0 \tag{14}$$

Here, the subscripts "$\perp$" (perp) and "||" (parallel) represent the components perpendicular and parallel to the background magnetic field, respectively. The simplified dispersion relation can be expressed as:

$$Ak_{\perp}^4 - Bk_{\perp}^2 + C = 0 \tag{15}$$

with

$$\begin{cases} A = S \\ B = \frac{\omega^2}{c^2}(S+P)(S - \frac{c^2}{\omega^2}k_{\|}^2) - \frac{\omega^2}{c^2}D^2 \\ C = \frac{\omega^4}{c^4}P[(S - \frac{c^2}{\omega^2}k_{\|}^2)^2 - D^2] \end{cases} \tag{16}$$

Using the root-finding formula, we can obtain:

$$k_{\perp\pm}^2 = \frac{B \pm \sqrt{B^2 - 4AC}}{2A} \tag{17}$$

From the positive and negative signs in the equation's solution, it can be seen that two distinct wave modes propagate in the plasma. When the positive sign is taken, the wave mode is referred to as the slow wave or TG wave; when the negative sign is taken, the wave mode is called the fast wave or helicon wave. The perpendicular wave vector $k_{\perp\pm}^2$ is a function of the plasma dielectric properties, which are determined by the density (n) and temperature ($T_\alpha$) of electrons and ions, the external magnetic field ($B$), the antenna frequency ($f$), and the wave vector component parallel to the magnetic field ($k_{\|}$). When the solution to the dispersion relation satisfies $k_{\perp+}^2 > 0$ or $k_{\perp-}^2 > 0$, the slow wave or fast wave can propagate in the plasma, respectively. Conversely, the slow wave or fast wave undergoes exponential attenuation and cannot propagate, a phenomenon known as "evanescence." The electromagnetic field radiated by an antenna can be decomposed into a superposition of waves with different parallel wave vectors. By calculating the power spectrum of the antenna using Fourier transform, the magnitude of the dominant wave vector of the electromagnetic waves excited by the antenna can be determined. In the Helimak, the power spectrum distribution of the antenna is obtained through pre-calculation in a seawater model. In this model, the plasma is described as a seawater medium [42] with a dielectric constant of 81 and an electrical conductivity of 4, and the calculation cross-section is set as an annulus, located 0.02 m in front of the antenna. Its calculation formula is as follows:

$$p_{\mathrm{rad}}(k_{\|}) = \frac{1}{8\pi^2}\int \mathrm{Re}(E_z(k_z, k_{\|}) \times H_{\|}^*(k_z, k_{\|}))dk_{\mathrm{z}} \tag{18}$$

Since the magnetic field direction in the Helimak model is toroidal direction, the subscripts "||" and "$z$" represent the toroidal direction and axial direction of the Helimak cylindrical coordinate system, respectively. Consequently, "$k_z$" and "$k_{\|}$" denote the z-component of the wave vector and the component of the wave vector parallel to the magnetic field, respectively.

Figure 5 compares the parallel power spectra of the four antennas, illustrating their distinct $k_{\|}$-space distributions and implications for helicon wave excitation. The RSA exhibits a pronounced peak at $k_{\|} \simeq 9\ \mathrm{m}^{-1}$, indicative of its capability to selectively drive short-wavelength helicon modes with strong parallel propagation, leading to high plasma absorption efficiency. In contrast, the SA shows a broader spectrum centered around $k_{\|} \simeq 7\ \mathrm{m}^{-1}$, suggesting moderate absorption to multiple parallel modes, which may enhance uniform plasma excitation but reduces peak power density per mode. The S-bend antenna has a much smaller main wave vector ($k_{\|} \simeq 0.6\ \mathrm{m}^{-1}$), and its spectrum exhibits a single-peak distribution. This results from its highly asymmetric current distribution. The comb antenna exhibits a nearly zero central wave number ($k_{\parallel} \sim 0.1\ \mathrm{m}^{-1}$ for numerical stability), corresponding to uniform parallel phasing and excitation of near-zero parallel modes, which can effectively drive radial plasma currents but is less efficient for launching propagating helicon waves. However, although the positions of the main peaks in the power spectra of the different antennas vary, their overall spectral widths are similar, and each peak exhibits a relatively large half-maximum width, indicating that these antennas share a similar spectral structure. The similarity in the power spectra suggests that although the spectral characteristics of each antenna differ, the electric fields they excite in the plasma are likely highly similar. Consequently, the variation in the absorption coefficient of each antenna is primarily governed by its specific geometry and the near-field distribution of the excited electric field.

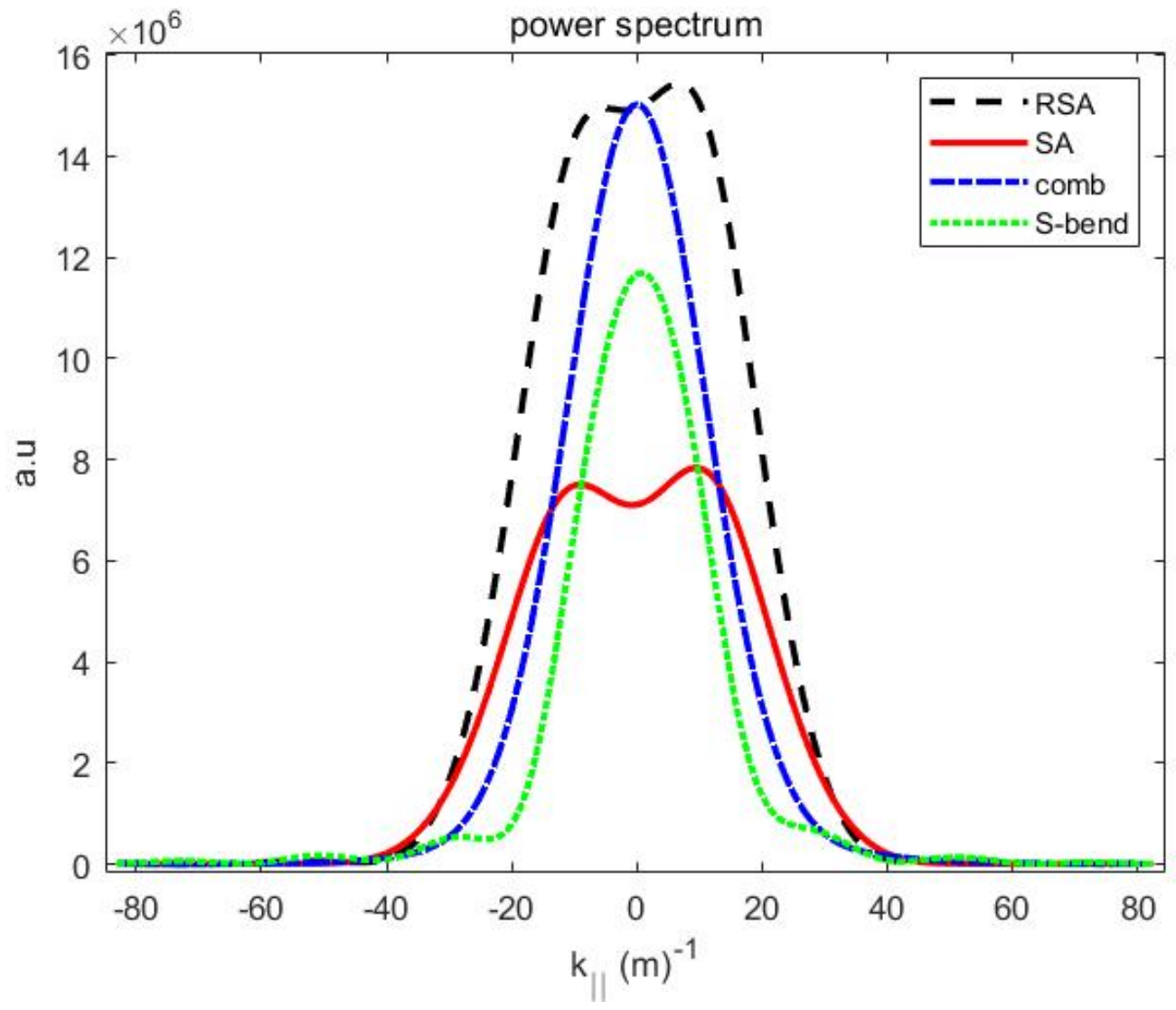

Figure 5. Power spectra of four antenna configurations: resonant strap antenna (RSA, black dashed), standard antenna (SA, red solid), comb antenna (blue dash-dotted), and S-bend antenna (green dotted).

Figure 6 presents the solutions for the perpendicular wavenumber of the helicon wave dispersion relation, corresponding to different antenna configurations, under the plasma parameters of the Helimak device. The results highlight the strong sensitivity of wave accessibility to both the plasma density profile and the choice of $k_{\parallel}$, revealing distinct propagation windows for the slow-wave and fast-wave branches. For the slow-wave branch, real solutions for $k_{\perp}$ exist only in the low-density region near the plasma boundary when $k_{\parallel}$ is sufficiently large (RSA, SA and S-bend antenna). This behaviour reflects the well-known requirement that slow waves must satisfy the electrostatic resonance condition $\omega \approx \omega_{\mathrm{pe}}$ in a magnetized plasma. At small $k_{\parallel}$ (comb antenna), the slow-wave perpendicular wavenumber becomes purely imaginary over the entire radial domain, indicating complete evanescence and explaining why slow waves generated by low-$k_{\parallel}$ antennas cannot penetrate beyond the edge SOL region. The fast-wave branch exhibits the opposite trend. At very small $k_{\parallel}$ (comb antenna), fast waves are permitted only in the high-density core, where the refractive index becomes sufficiently large to prevent mode conversion or cutoff. As $k_{\parallel}$ increases toward approximately $k_{\parallel} \approx 0.6\ \mathrm{m}^{-1}$ (S-bend antenna), the entire plasma column becomes accessible to fast-wave propagation. This regime corresponds to the classic "low-$n_{\parallel}$" helicon wave condition, where the wave fronts are nearly parallel to the magnetic field and the dielectric tensor favours whistler-like propagation. Beyond this optimum value, further increases in $k_{\parallel}$ (RSA and SA) again lead to global cutoff of the fast wave across the full radius, as the dispersion relation transitions toward the high-$n_{\parallel}$ electrostatic limit. Taken together, the dispersion characteristics in Figure 6 explain the experimentally observed difficulty of depositing helicon wave power into the Helimak core. The antennas used in the present configuration predominantly excite either excessively small $k_\parallel$ (for the S-bend and comb antennas) or excessively large $k_\parallel$ (for the spiral antennas), placing their dominant spectral content outside the narrow accessibility windows identified by the dispersion analysis.

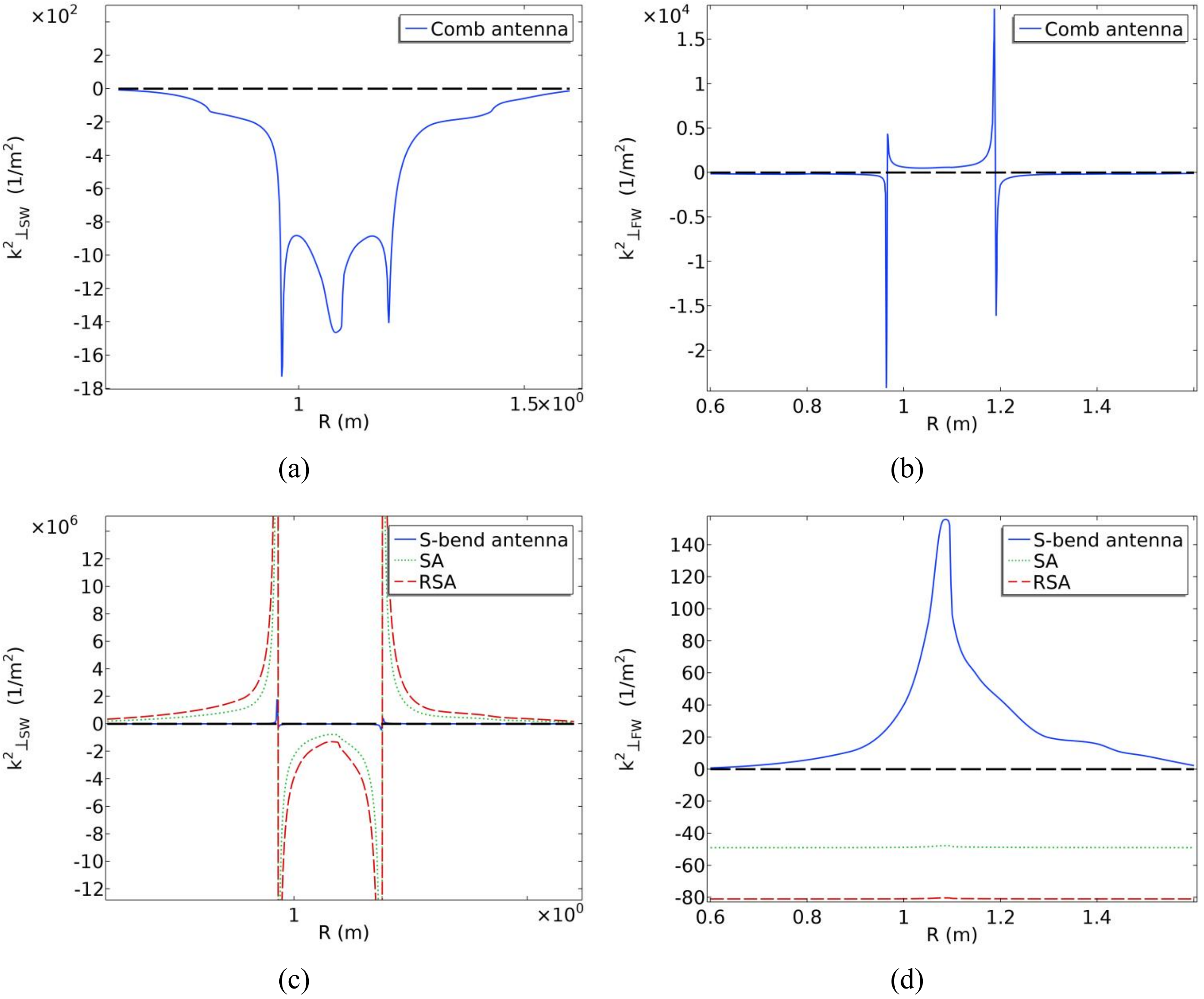

Figure 6. Radial wave-number distributions in the Helimak configuration: (a) Trivelpiece–Gould (TG) wave number excited by the comb antenna; (b) helicon wave number excited by the comb antenna; (c) TG wave numbers excited by the resonant strap antenna (RSA), standard antenna (SA), and S-bend antenna; (d) corresponding helicon wave numbers for the RSA, SA, and S-bend antennas.

### *2.5 Justification and validity of plasma model assumptions*

The plasma model employed in this study adopts several simplifying assumptions that are appropriate for the Helimak operating regime but differ in some respects from the tokamak edge plasma. These assumptions are introduced to isolate the dominant physics governing power absorption under low-temperature, low-density, electron-heated conditions, and their validity is summarized in Table 1 and discussed below.

The electron and ion temperatures are assumed to be spatially uniform and constant in time. This approximation is supported by experimental measurements in Helimak, which indicate relatively weak temperature gradients across the radial region of interest and electron temperatures typically in the range $T_e \sim 5 - 15$ eV. In this regime, RF power deposition primarily modifies plasma density rather than temperature, and temperature fluctuations are second-order effects for the absorbed metrics considered here. To assess sensitivity, additional simulations were performed varying $T_e$ between 5 and 15 eV, which did not alter the qualitative dependence on antenna geometry. This confirms that the principal conclusions are robust against moderate temperature variations within the experimentally relevant range.

Quasi-neutrality is enforced by assuming equal electron and ion densities. Given the characteristic spatial scales of the RF fields considered here, which are much larger than the Debye length under Helimak conditions, charge separation effects are negligible and the quasi-neutral approximation is well justified. This assumption is standard in low-frequency RF modelling of open-field-line plasmas and does not affect the evaluation of plasma absorption efficiency.

Plasma profiles are assumed to vary only in the radial direction, while poloidal and toroidal variations are neglected. This approximation reflects the dominant radial gradients observed experimentally in Helimak and allows a clear interpretation of the role of antenna near-field structure in determining absorption efficiency. While 3D effects may influence local field distributions, they are not expected to qualitatively modify the parametric trends reported in this work.

Finally, a constant effective collision frequency dominated by electron–neutral collisions is assumed. This reflects the relatively high neutral pressure and low electron temperature in Helimak, for which electron–neutral collisions exceed electron–ion collisions across most of the plasma column. Variations in collision frequency primarily affect absolute damping rates but do not significantly change the relative power absorbed trends between different antenna configurations.

Table 1. Summary of plasma model assumptions, validity range, and expected impact on RF absorption.

| **Assumption** | **Physical justification (Helimak)** | **Validity range** | **Expected impact on results** |
|---|---|---|---|
| Constant $T_e$ and $T_i$ | Weak temperature gradients; electron-heated plasma with $T_e \sim 5 - 15$eV | Low-temperature, steady-state Helimak discharges | Affects absolute damping; qualitative absorption trends unchanged |
| $n_e = n_i$ (quasi-neutrality) | RF spatial scales $\gg \lambda_D$ | All conditions studied | Negligible impact on absorption efficiency |
| Radial-only profile variation | Dominant experimental gradients are radial | Axisymmetric Helimak operation | Simplifies interpretation; trends preserved |
| Constant collision frequency | Electron–neutral collisions dominate | High neutral pressure, low $T_e$ | Sets damping level; relative comparisons unaffected |

### *2.6 Supplementary Explanation of Absorption Efficiency*

The power absorbed by the plasma is evaluated by integrating the local power dissipation over a defined volume. To avoid ambiguity, two distinct integration volumes are clearly defined for the first time in this work: the "entire plasma volume" refers to the full computational domain containing finite plasma density, including both the vacuum chamber and the waveguide region; whereas the "main plasma volume" is defined as the radially confined vacuum chamber region. Moreover, due to the linear nature of Maxwell equations, the magnitude of the forward power ($P_{\text{fwd}}$) in the transmission line does not alter the impedance characteristics of the system—that is, the power absorption efficiency of the plasma in Helimak does not vary with $P_{\text{fwd}}$. Hence, the subsequent analysis in this work employs absorption efficiency rather than absorbed power. Although even with a low-absorption-coefficient antenna, good impedance matching can enable the radiation of power comparable to that of a high-efficiency antenna and effectively alter the plasma density distribution in the Helimak device, designing and employing antennas with a higher absorption coefficient remains of significant value. A higher absorption coefficient corresponds to a lower reflection coefficient at the antenna port and a lower standing wave voltage in the transmission line. Consequently, for a given transmission line breakdown voltage, a larger absorption coefficient leads to a higher operational power limit for the antenna.

## 3. Results and Analyses

### *3.1. Helicon Wave Heating Under Current Window*

Using THEMIS, we computed the helicon wave heating for the Helimak device under current (this section) and optimized (next section) windows, respectively. Figure 7 illustrates the radial electric field distributions excited by the four types of antennas in the Helimak plasma. The simulation results indicate that the radial component dominates the electric field in all cases. The volume-averaged values of proportion of the radial electric field in the total electric field for RSA, SA, comb antenna, and S-bend antenna are calculated to be 84.6%, 90.2%, 87.1%, and 84.7%, respectively. The electric field structure clearly exhibits wave mode characteristics propagating radially, meaning the electric field direction is found to be essentially parallel to the wave vector direction. This result is consistent with the quasi-electrostatic nature of slow waves and agrees with theoretical predictions from the dispersion relation. Figure 8 presents the power deposition distributions for the four antenna types, revealing that energy deposition is consistently concentrated near the ceramic window and the adjacent metal wall regions. The deposition distributions display a clear standing-wave pattern, which is likely caused by the radial gradient in electron density. This gradient results in partial reflection of the incident wave, and the interference between the reflected and incoming waves leads to the formation of a standing-wave structure. Table 2 presents the contribution ratios of four heating mechanisms to the total power deposition for each of the four antenna types. Since the simulation results indicate that the ion contribution to the total power deposition is below $1\times10^{-5}$ and thus negligible, only the electron contribution is provided in the table. A comparison of the relative contributions across different antennas reveals that the power deposition shares of the four mechanisms remain essentially consistent, indicating that the shares of heating mechanisms are independent of the antenna geometry. Namely, geometry mainly affects absorption, not damping partition, under these conditions. Among them, Landau damping is the dominant power deposition mechanism, accounting for approximately 68% of the total, followed by collisional damping at about 30%. The contributions from Doppler-shifted cyclotron damping and anomalous Doppler damping are nearly identical, lying in 0.3%- 0.4%, and are considered insignificant.

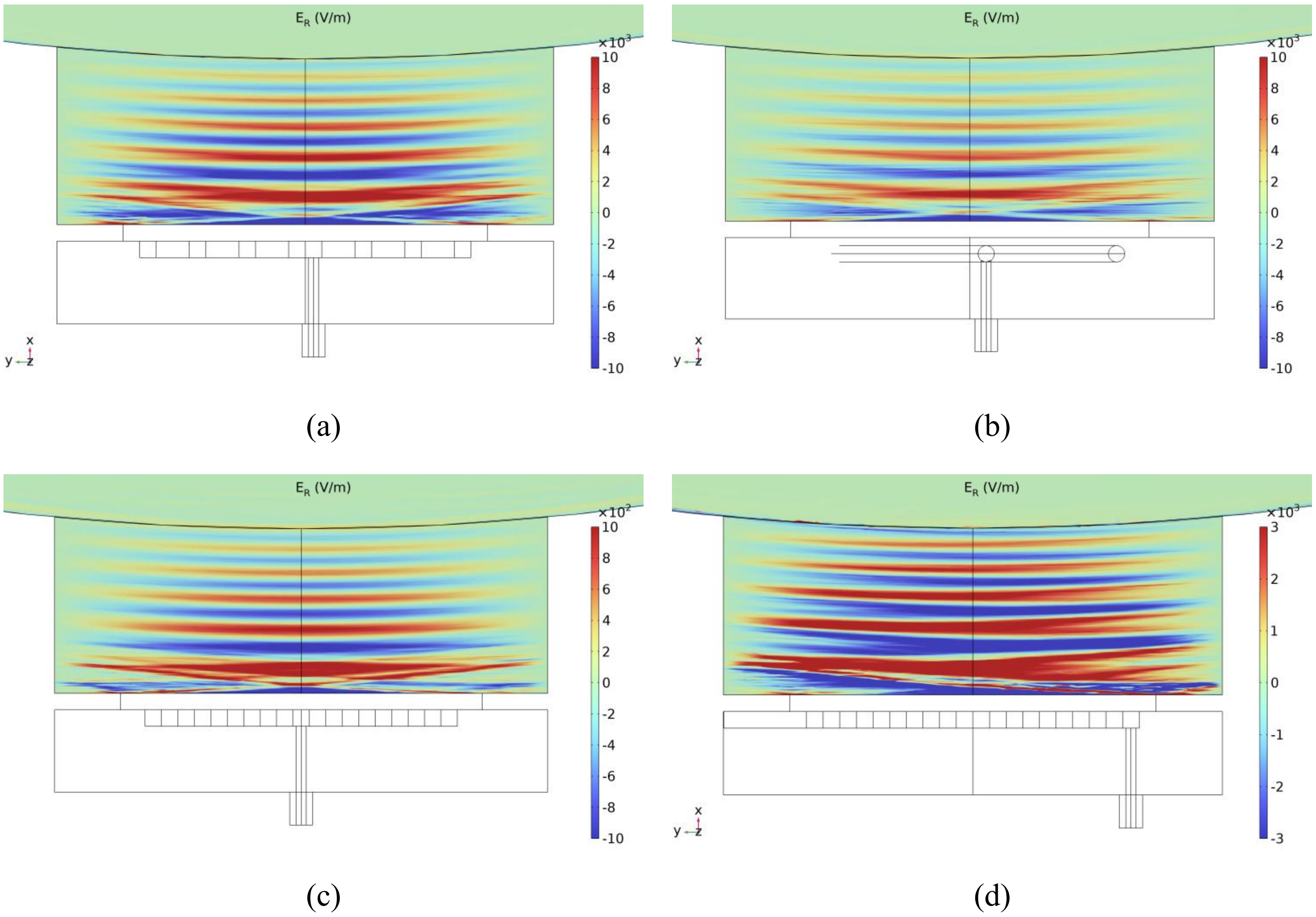

Figure 7. Radial profiles of the antenna-driven electric field in the Helimak geometry for (a) the resonant strap antenna (RSA), (b) the standard antenna (SA), (c) the comb antenna, and (d) the S-bend antenna, illustrating differences in radial field localization.

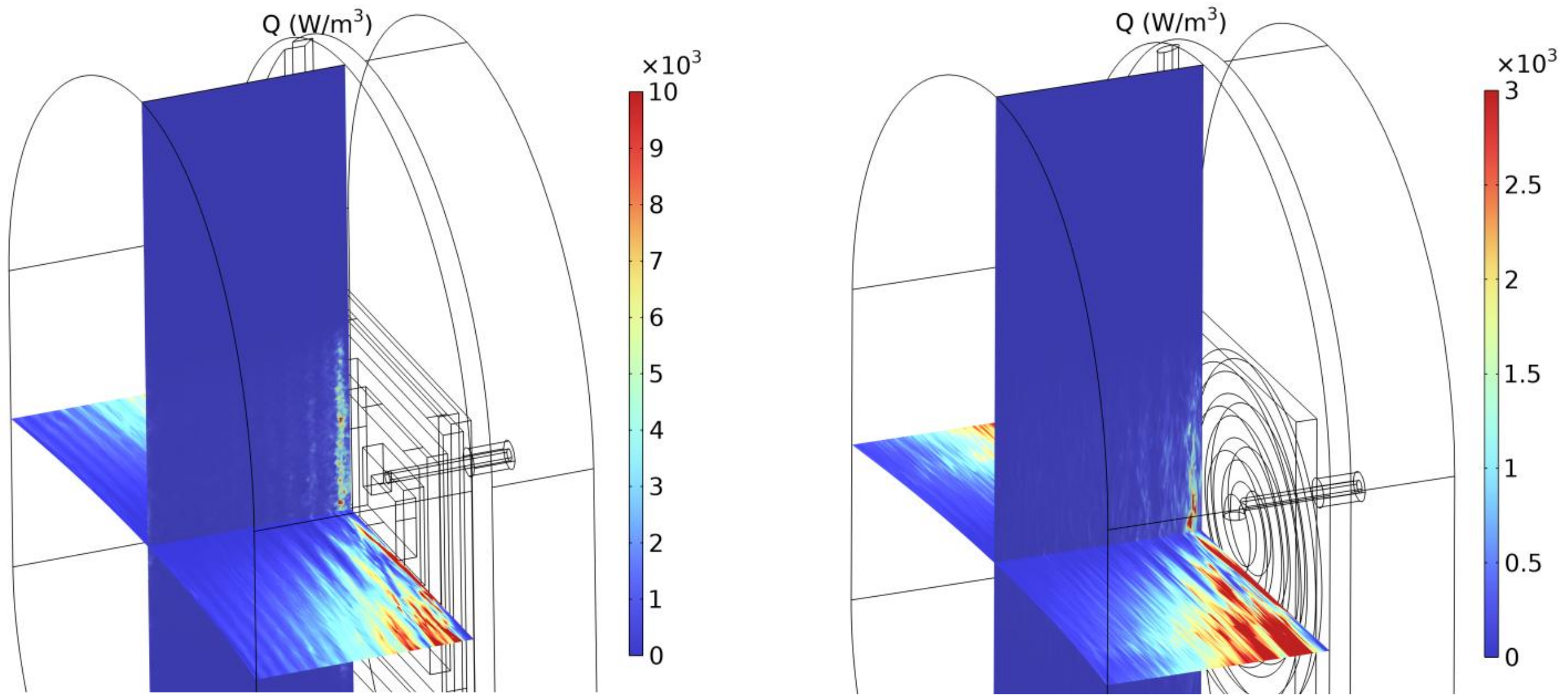

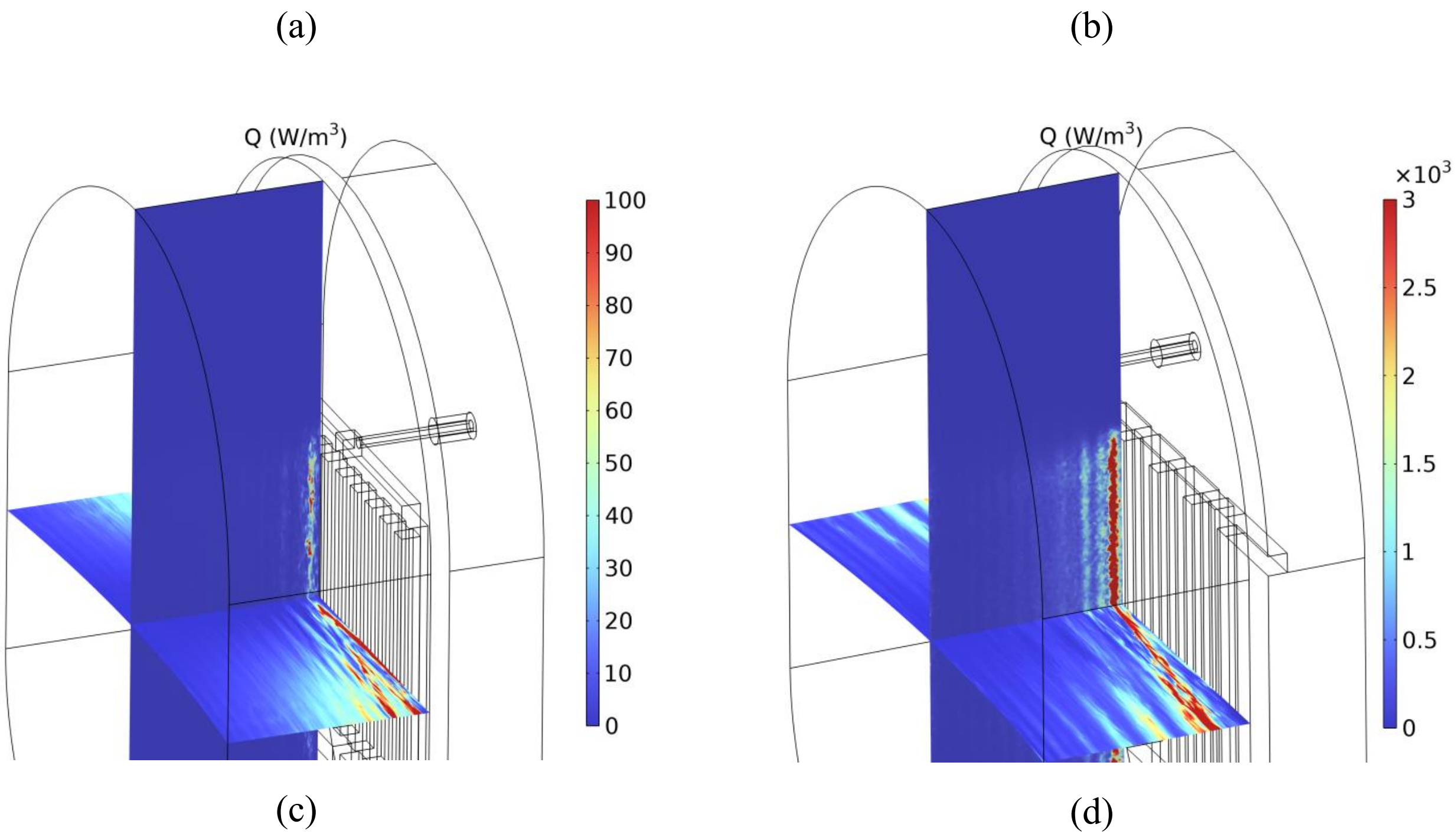

Figure 8. Radial distributions of the power deposition for the four antenna types, illustrating differences in localization: (a) rectangular spiral antenna (RSA), (b) spiral antenna (SA), (c) comb antenna, and (d) S-bend antenna.

Table 2. Relative proportions of the four electron power-deposition mechanisms for each of the four antenna configurations investigated.

| **Antenna Type** | **DSCD** | **ADD** | **CD** | **LD** |
|---|---|---|---|---|
| RSA | 0.39% | 0.40% | 31% | 67% |
| SA | 0.37% | 0.38% | 30% | 68% |
| Comb antenna | 0.37% | 0.38% | 31% | 67% |
| S-bend antenna | 0.35% | 0.36% | 33% | 66% |

The fact that the contribution ratios of the different heating mechanisms are invariant with antenna geometry stems from a key physical constraint: the electron cyclotron frequency greatly exceeds the parallel phase velocity $\omega_{ce} \gg k_{\parallel} v_{\mathrm{T}e}$. Under this condition, the plasma dispersion function $Z\left(\zeta_{\pm 1}^{e}\right)$ is effectively zero, suppressing damping mechanisms associated with the wave's circularly polarized components. Therefore, power deposition is dominated solely by the component of the electric field parallel to the magnetic field. According to Equation (9), the ratio of power deposition from electron collisional damping to that from electron Landau damping is given by:

$$\frac{P_{\text{abs_CD_e}}}{P_{\text{abs_LD_e}}}=\frac{\frac{1}{2}\omega\varepsilon_0\cdot\operatorname{Im}\left\{1-\frac{\omega_{\text{pe}}^2}{\omega\left(\omega+j\nu_{\text{e}}\right)}\right\}\cdot\left|E_{\parallel}\right|^2}{\frac{1}{2}\omega\varepsilon_0\cdot\operatorname{Im}\left\{1-\frac{\omega_{\text{pe}}^2}{k_{\text{z}}^2\upsilon_{\text{Te}}^2}\cdot Z'\left(\zeta_0^{\text{e}}\right)\right\}\cdot\left|E_{\parallel}\right|^2}=\operatorname{Im}\left\{\frac{k_{\text{z}}^2\upsilon_{\text{Te}}^2}{\omega\left(\omega+j\nu_{\text{e}}\right)Z'\left(\frac{\omega}{\left|k_{\text{z}}\right|\upsilon_{\text{Te}}}\right)}\right\} \tag{19}$$

The source frequency and plasma temperature in Equation (19) are set as constants in the model. The electron collision frequency, dominated by electron-neutral collisions, is also treated as constant due to the fixed background neutral pressure. Although the four antennas exhibit different parallel wave numbers, the Landau damping term remains independent of parallel wave number under low electron temperature condition. As a result, the ratio of power deposition contributed by electron collisional damping to that by electron Landau damping is essentially consistent across all four antenna types. Despite variations in their relative performance (RSA > SA > S-bend > comb), all four antennas exhibit prohibitively low absorption efficiency of main plasma under the current window configuration, as shown in Table 3.

Table 3. Comparison of the absorption efficiencies for the four antenna configurations investigated.

| **Antenna Type** | **absorption Efficiency (entire plasma)** | **absorption Efficiency (main plasma)** |
|---|---|---|
| RSA | 7.54 % | 0.14 % |
| SA | 3.08 % | 0.050% |
| Comb antenna | 0.097% | 0.0027% |
| S-bend antenna | 2.74 % | 0.084% |

The simulated power deposition is confined almost entirely to the wave guide channel, a finding corroborated by experimental probe data showing no measurable effect on the core plasma density. Simulation and experimental results reveal multiple physical and engineering challenges: (i) the slow wave exhibits rapid attenuation in the low-density region of the scrape-off layer, indicating an inherent limitation in its penetration capability; (ii) the waveguide channel formed by the current protruding window structure further restricts the transmission of wave energy toward the main plasma; (iii) under the current conditions, the antenna design has a high reflection coefficient, which increases the practical challenge of achieving impedance matching and poses difficulties in efficient power injection. These factors collectively lead to ineffective control of the main plasma via helicon waves under the existing window configuration.

Therefore, numerical simulations indicate that the wave electric field is predominantly radial, in agreement with the characteristics of slow-wave propagation. The power absorbed exhibits a pronounced standing-wave structure localized near the ceramic window, arising from strong wave reflections in the low-density edge plasma. Analysis of the dissipation channels shows that Landau damping accounts for approximately 68% of the total power absorption, while collisional damping contributes about 30%; in contrast, cyclotron-related mechanisms provide a negligible contribution (<0.4%). These ratios are essentially insensitive to the geometry of the antenna. Consistent with experimental observations of no measurable density increase, the absorption efficiency of RF power for the core plasma remains extremely weak, below 0.2%. This poor absorption can be attributed to three primary factors: the slow-wave cutoff in low-density regions, which inhibits wave propagation toward the plasma core; wave trapping within the protruding ceramic window that limits radial penetration; and the high reflection coefficients at the plasma–antenna interface, which complicate impedance matching and severely restrict effective power injection. These findings indicate that the current window configuration is unsuitable for helicon-assisted boundary-density control.

### *3.2. Helicon Wave Heating Under New Recessed Window*

To address the technical challenge of low absorption efficiency, an optimized window design featuring a recessed configuration has been developed. The new design incorporates a concave flange that will position the ceramic window inside the main vacuum chamber, enabling direct, close-range heating of the plasma. Furthermore, the window shape has been redesigned from a rectangle to a racetrack to increase the effective radiating area, which is expected to improve wave absorption efficiency. As shown in Figure 9, benefiting from the enlarged ceramic window area of the new design, this study expands the test range of antenna layout schemes and includes simulation results after a 90° rotation of the antenna for comparison.

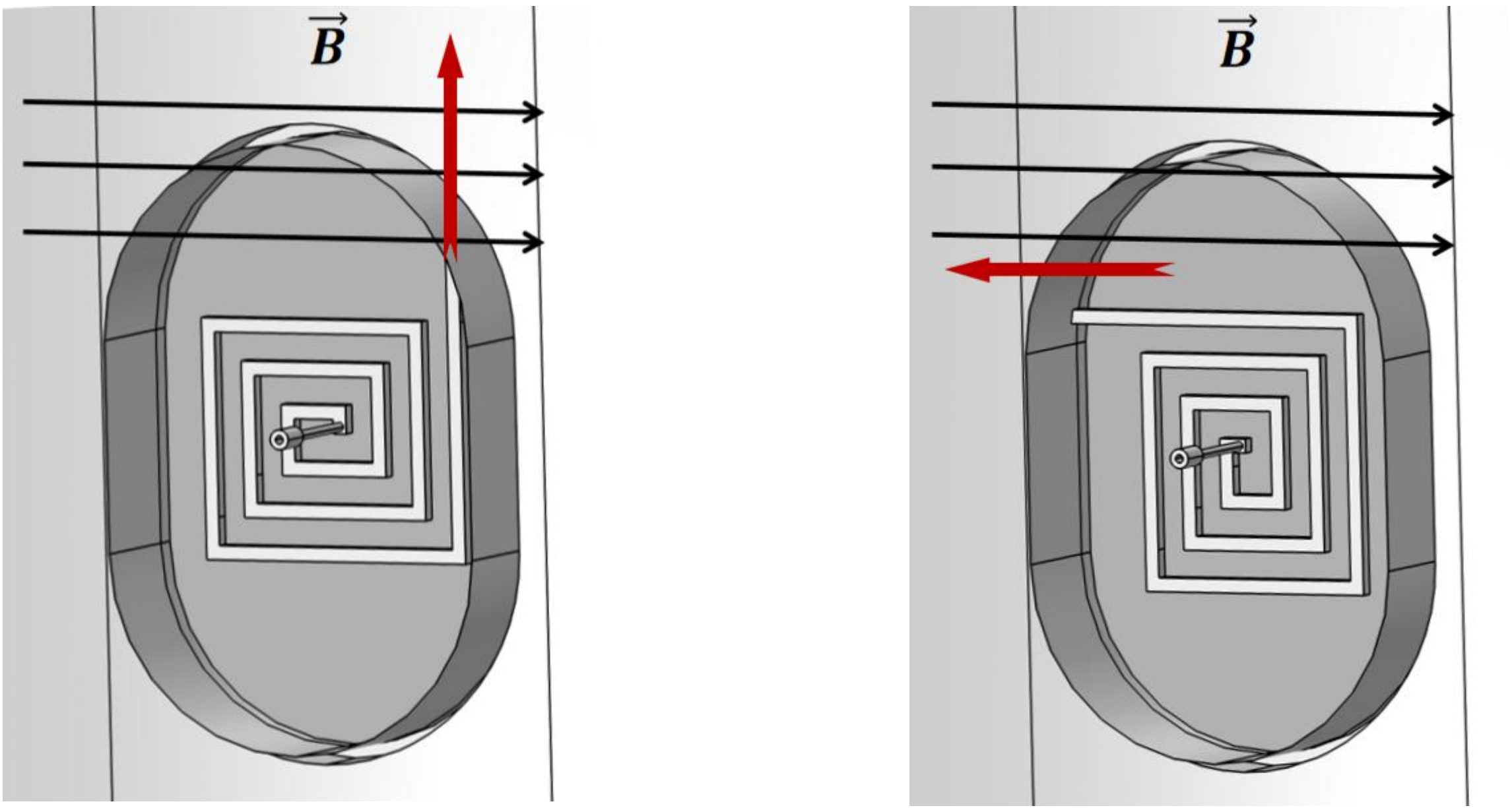

(a) (b)

Figure 9. New recessed window and helicon wave antenna: (a) current-strips oriented perpendicular to the magnetic field; (b) current-strips oriented parallel to the magnetic field.

Based on the angular relationship between the direction of the current-strip at the antenna end and the background magnetic field direction, the antenna layouts are mainly divided into two types: one with the end current-strip perpendicular to the magnetic field, and the other with it parallel to the magnetic field. In the previous section, an iterative mesh refinement was applied to the Helimak model for electromagnetic wave propagation simulations, focusing on key areas such as the waveguide region and the vicinity of the vacuum chamber wall. In this section, we focuses on the recessed window configuration. Since the wave propagation region is significantly expanded, to control the computational scale, the simulation domain is reduced to cover a quarter of the vacuum chamber spanning 90° in the toroidal direction. The cross-sections at both ends of the vacuum chamber adopt periodic boundary conditions.

As shown in Figures 10, under the new recessed window configurations with different recess distances, the ranking of the absorption efficiency for the four antenna types (from highest to lowest) remains largely consistent with that under the original window condition. Among them, RSA continues to exhibit the best power absorption performance, while SA shows a absorption efficiency comparable to that of the S-bend antenna. The comb antenna demonstrates the lowest absorption efficiency. Moreover, the sensitivity of each antenna to its placement orientation depends on its specific geometric characteristics. The comb antenna, being less symmetric, is the most affected by orientation changes. In contrast, the rectangular spiral and spiral antennas exhibit greater consistency in their geometric layout along certain directions, resulting in lower sensitivity. Overall, the influence of placement orientation on absorption efficiency remains limited. Additionally, as illustrated in Figures 10, we systematically examine the impact of window's radial position and antenna's orientation on power absorption within the new recessed window structure through parametric scanning. Simulation results indicate that antenna absorption efficiency exhibits non-monotonic relationship with variations in the radial position of the window.

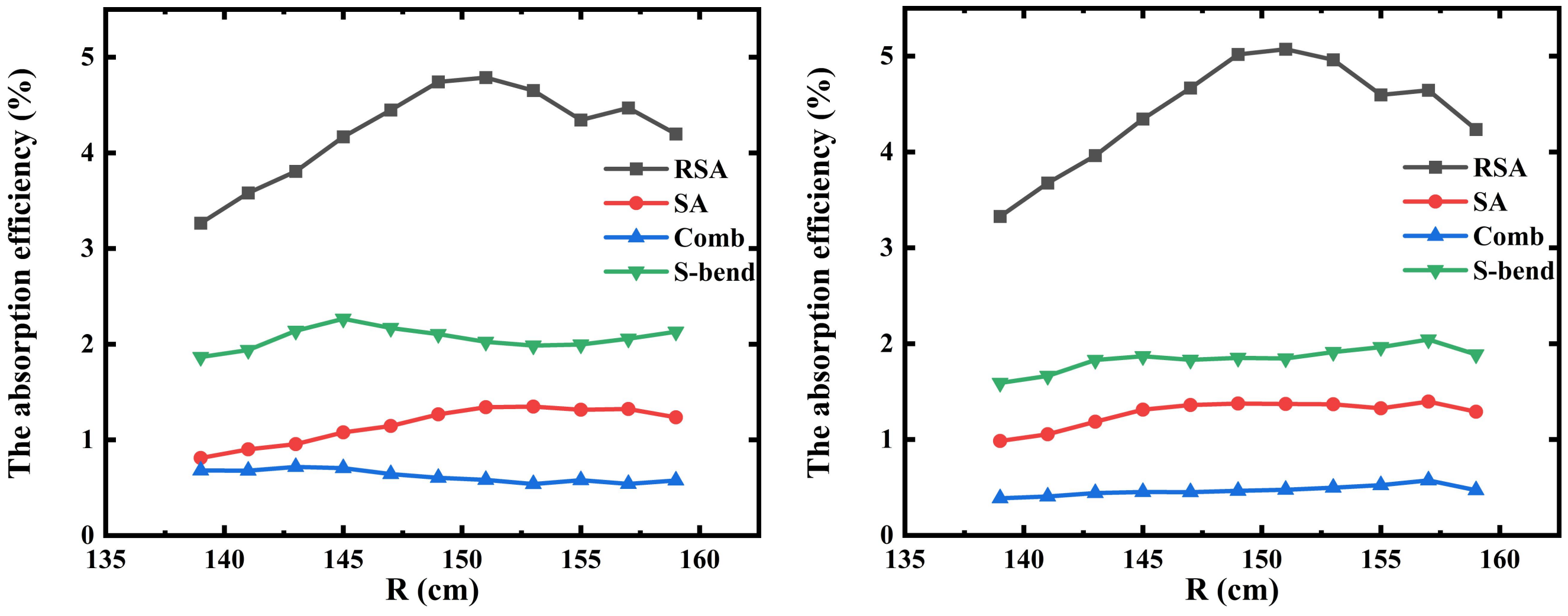

Figure 10. Dependence of the absorption efficiency of the four antenna configurations on the radial position of the window: (a) current-strips oriented perpendicular to the magnetic field; (b) current-strips oriented parallel to the magnetic field.

Specifically, the absorption efficiency of the rectangular spiral antenna and the spiral antenna shows a double-peak characteristic as the radial position of the window changes, with peaks occurring near major radii of approximately 0.153 m and 0.157 m, respectively. In contrast, the response of the comb antenna and the S-bend antenna differs. Taking the S-bend antenna as an example, when the end of the current-strip is perpendicular to the magnetic field, the absorption efficiency exhibits only a single peak near a major radius of about 0.145 m; when the end of the current-strip is parallel to the magnetic field, in addition to the peak at 0.145 m, an additional peak appears near 0.157 m. This non-monotonic variation suggests that specific optimal positions exist in the recessed window design, and local maximization of absorption efficiency can be achieved by precisely controlling the radial position of the window. However, the optimal positions corresponding to each peak vary depending on the antenna type and orientation, reflecting the complexity of the interaction between the boundary plasma and the near-field of the antenna.

### *3.3. Optimization of Spiral Antennas Under New Recessed Window*

Based on a comparison of the absorption efficiency of the four antenna types, we consider that spiral-type antennas (including RSA and SA) exhibit greater potential for optimization. To further explore the design principles of antennas with higher absorption efficiency, we conducted a parametric study on the key structural parameters of spiral-type antennas. As shown in Figure 11, the reference parameters of RSA are: conductor strap width of 0.01 m, inter-turn spacing of 0.02 m, and 2.5 turns. The reference structural parameters of SA are as follows: conductor strip diameter of 0.01 m, inter-turn spacing of 0.01 m, and 4 turns. In this section, a systematic parameter scan will be conducted focusing on three key aspects: the grounding configuration of the current-strip termination, the antenna box dimensions, and the geometric parameters of the current-strip. This parameter-scan study aims to establish optimized design guidelines for helical antennas, elucidate the underlying physical mechanisms, and thereby deepen the understanding of the absorption mechanism between the antenna and boundary plasma.

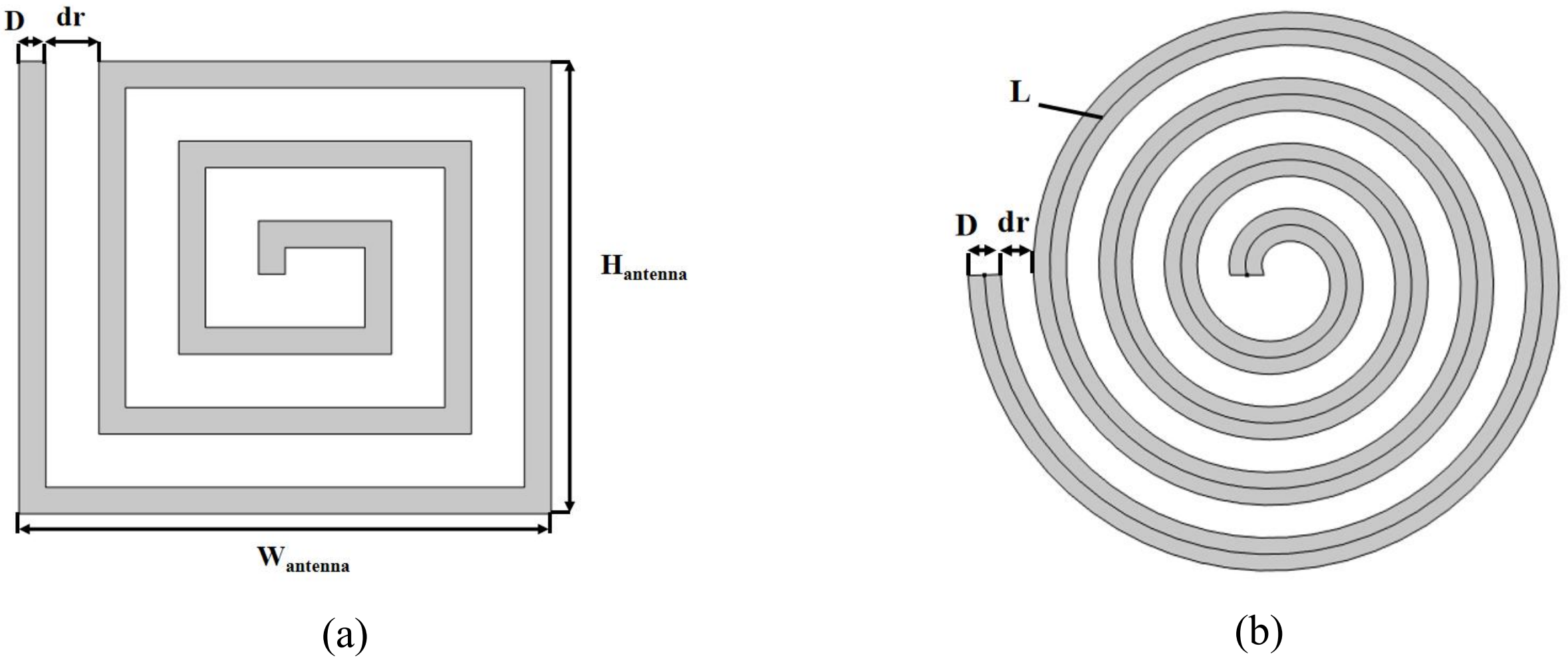

Figure 11. Parametric structure diagrams of the investigated antenna types: (a) RSA; (b) SA.

### 3.3.1 Comparison of short-circuit and open-circuit terminals

Based on transmission line theory, the design of the antenna current-strip can be divided into two terminal strategies: open-circuit and short-circuit. The selection of them significantly affects aspects such as the antenna's coupled power and stability. The short-circuit terminal design directly connects the inner conductor and outer conductor of the coaxial transmission line at the end, forming a low-impedance point where a current antinode and a voltage node occur. This characteristic enables it to withstand extremely high RF currents and significantly reduces the risk of electrical breakdown. Meanwhile, the short-circuit design also features greater mechanical strength and a more efficient heat conduction path, making it easier to achieve cooling and thermal management under megawatt-level power. Therefore, short-circuit designs are commonly adopted for high-power antennas in tokamak devices. In contrast, the open-circuit terminal design leaves the conductor end suspended, forming a high-impedance point where a voltage antinode and a current node exist. Although its structure is relatively simple, the high electric field intensity at its terminal can easily trigger plasma or vacuum breakdown. Since megawatt-level high-power sources are not required in boundary helicon wave discharge heating experiments, the simplicity and high electric field intensity of the open-circuit design may make it an effective power absorption design scheme.

Table 4 compares the absorption efficiency of RSA and SA under both short-circuit and open-circuit terminal configurations, with the new window recessed 0.05 m into the chamber. The data show that the open-circuit termination significantly enhances the absorption efficiency of both antenna types—by roughly a factor of 3–4—relative to the short-circuit design. This improvement arises from the high-impedance boundary condition at the open-circuit terminal, which reduces reflection at the antenna-plasma interface and facilitates more effective helicon wave excitation. In particular, the RSA configuration consistently exhibits higher absorption efficiency than the SA configuration due to its larger effective current-strip area. These results highlight the critical role of terminal impedance and antenna geometry in optimizing power absorption.

Table 4. Comparison of the absorption efficiency of RSA and SA antennas under short-circuit and open-circuit terminal configurations, with the window recessed 0.05 m into the chamber.

| **Terminal configuration** | **RSA** | **SA** |
|---|---|---|
| Short circuit | 4.4 % | 1.3 % |
| Open circuit | 17 % | 5.7 % |

Figure 12 further compares the radial electric field distribution of SA under the two terminal designs. The results indicate that the open-circuit termination substantially modifies the electric field near the window–plasma interface, increasing field amplitude and promoting stronger local power absorption. In contrast, the radial field distribution within the main plasma column remains largely unchanged, suggesting that the overall wave propagation and helicon mode structure are preserved. Based on the above observations, for the Helimak device, the open-circuit design of the current strap termination primarily enhances absorption efficiency by optimizing the electric field distribution at the plasma edge. This approach provides an effective strategy for maximizing antenna power transfer without significantly altering the overall wave dynamics.

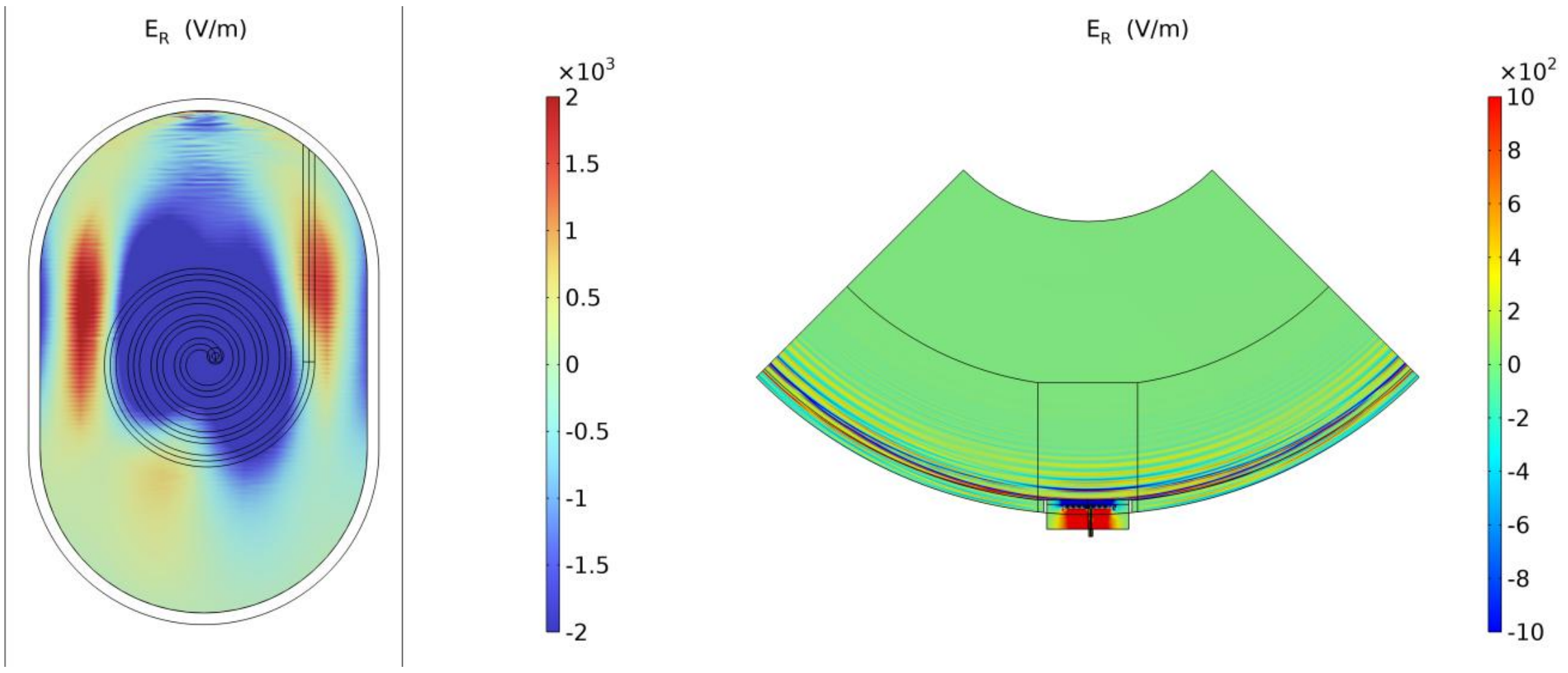

(a) (b)

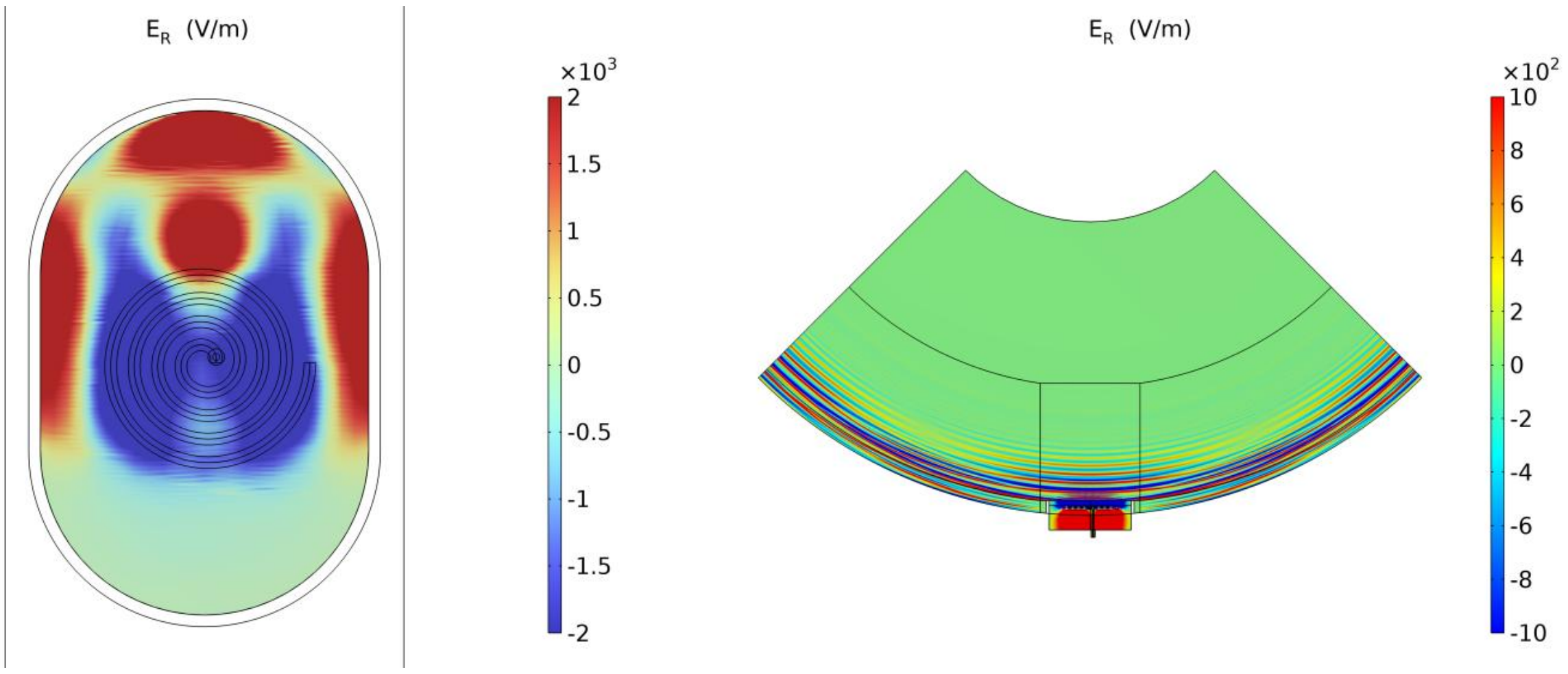

(c) (d)

Figure 12. Radial distribution of the electric field for the SA antenna under short-circuit and open-circuit terminal configurations. Panels show (a) the window–plasma interface under short-circuit conditions, (b) the antenna mid-plane under short-circuit conditions, (c) the window–plasma interface under open-circuit conditions, and (d) the antenna mid-plane under open-circuit conditions.

### 3.3.2 Influence of antenna box shape on absorption efficiency

In the design of a helicon wave antenna system, the antenna box serves not merely as a metallic enclosure or protective housing, but as a critical functional component within the energy absorption system. Its rational structural design and optimization play an important role in achieving efficient and stable helicon wave plasma heating. Currently, the geometry and overall dimensions of the antenna box are constrained by the window flange structure and have been fixed as a racetrack configuration. Therefore, this section focuses on a parametric scan of the width and depth parameters of the antenna box for the open-circuit helicon wave antenna. Figure 13 illustrates the correspondence between the absorption efficiency and the variation in the width and depth of the antenna box. The parametric study demonstrates a clear trend: the proximity of the box walls imposes a limiting effect on the absorption efficiency. As the distance between the antenna and the box wall decreases, the absorption efficiency systematically decreases due to enhanced electromagnetic reflection and confinement effects that restrict the radiated wave from effectively penetrating the plasma. Conversely, increasing the box dimensions mitigates these boundary effects, allowing for stronger wave–plasma interaction and improved power transfer. These results indicate that, within mechanical constraints, maximizing the antenna box width and depth is critical for achieving optimal helicon wave absorption and overall antenna performance.

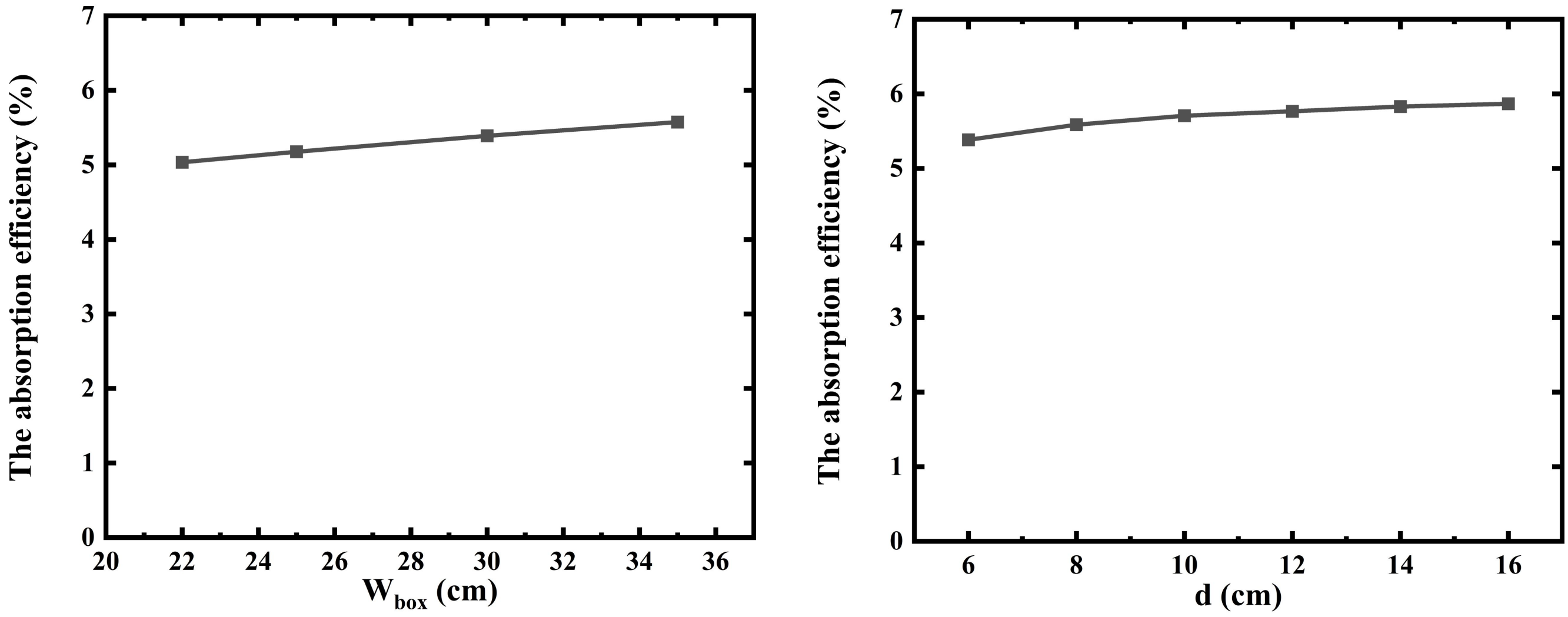

(a) (b)

Figure 13. Influence of antenna box dimensions on the absorption efficiency: (a) variation with box width; (b) variation with box depth.

3.3.3 Effect of current-strip length on absorption efficiency

When the number of turns of the spiral antenna is 2, 3, 4, and 5 respectively, the lengths of the current-strip are approximately 0.38 m, 0.76 m, 1.26 m, and 1.89 m. Figure 14 shows the relationship between the absorption efficiency of the spiral antenna and the length of the current-strip. The results show that the absorption efficiency exhibits an approximately linear increase with the length of the current-strip, which may be related to the linear expansion of the radiating area with the strip length. Owing to the anisotropic nature of magnetized plasma, the orientation of the current-strip relative to the background magnetic field significantly influences its radiation characteristics. To quantitatively investigate the effects of the current-strip lengths perpendicular to and parallel to the magnetic field on the absorption efficiency, a systematic parametric scan of the width and length of RSA was conducted in this study. Figure 15 presents the results of a parametric scan on two key dimensions of a rectangular spiral antenna: the antenna width along the magnetic field direction and the height perpendicular to the magnetic field direction. The results demonstrate that variations in the antenna height (dimension perpendicular to the magnetic field) have a considerably greater impact on absorption efficiency than variations in the antenna width (dimension parallel to the magnetic field). This phenomenon can be explained from the perspectives of wave propagation characteristics and the antenna near-field excitation mechanism: in plasma regions where slow-wave propagation dominates, the electric field diffuse efficiently along the magnetic field direction. As a result, changes in the conductor length parallel to the magnetic field have limited influence on the field distribution. In contrast, extending the strap perpendicular to the magnetic field effectively enlarges the radiation area and the propagation range of the wave, thereby significantly enhancing power absorption. This finding provides clear guidance for optimizing antenna design aimed at efficient absorption with magnetized plasmas. Furthermore, in Figure 15(a), when the antenna width exceeds 23 cm, the growth rate of absorption efficiency is significantly suppressed. This may be because the width of the window along the direction parallel to the magnetic field is limited; as the antenna width increases, the current-strip becomes closer to the antenna box wall, leading to the suppression of absorption efficiency by the antenna box wall.

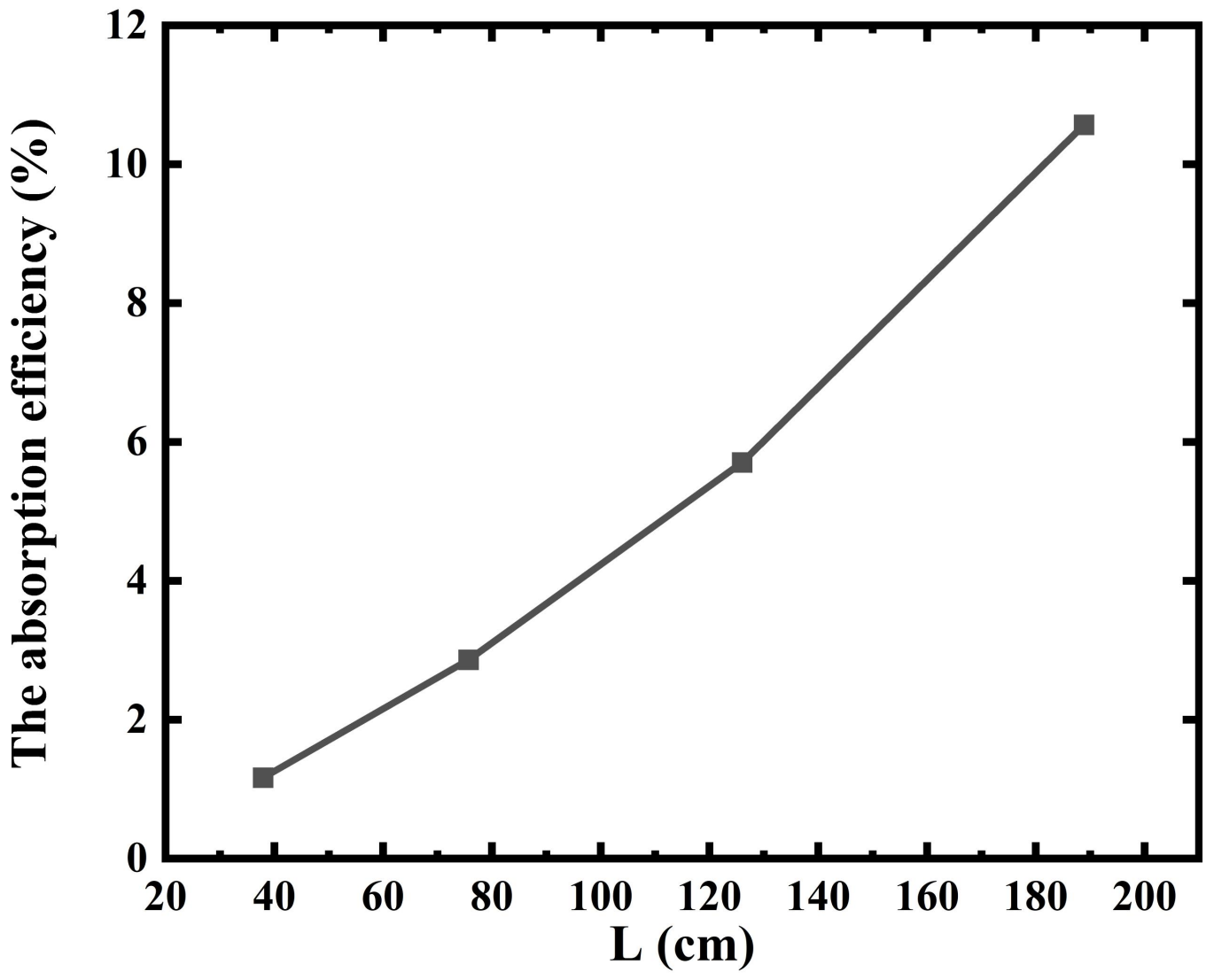

Figure 14. Dependence of the SA antenna absorption efficiency on the current-strip length.

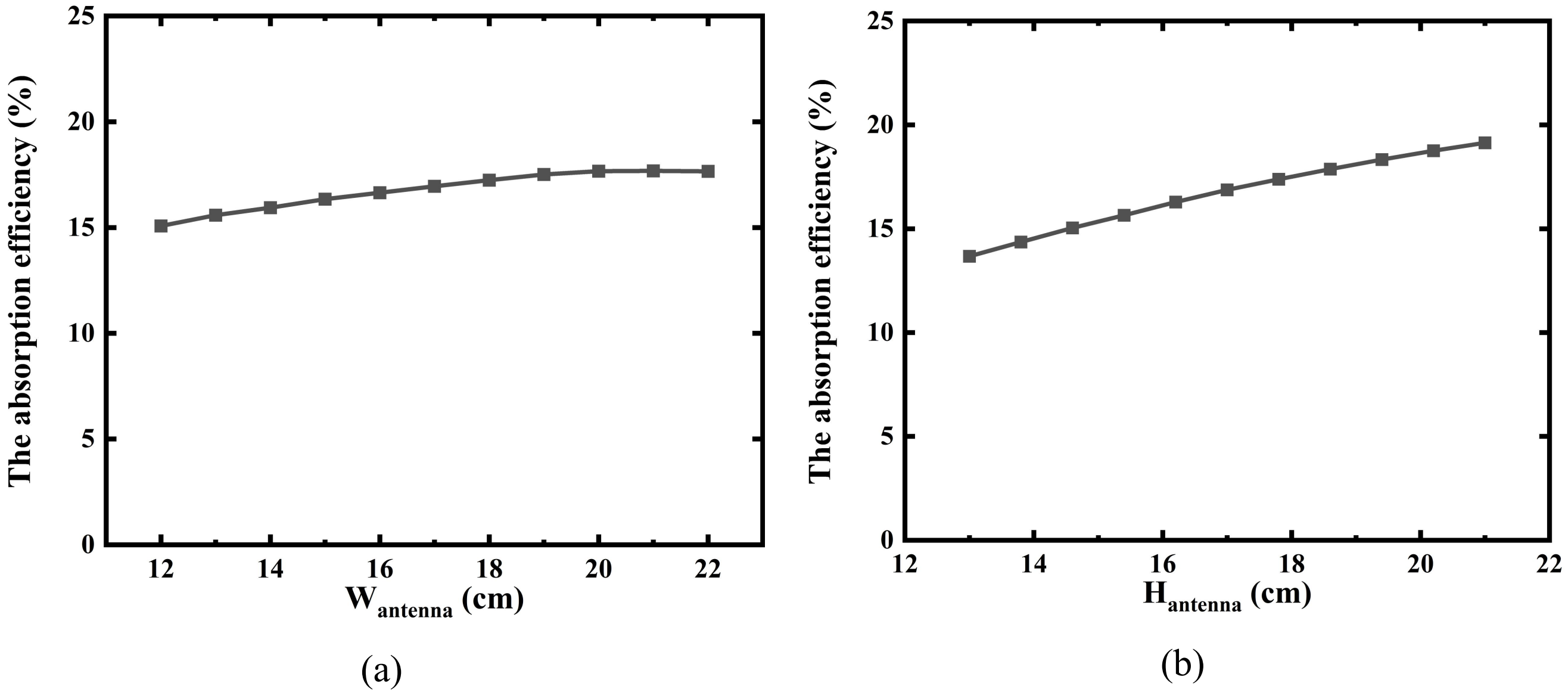

Figure 15. Effect of antenna size on the absorption efficiency of the RSA: (a) variation with antenna width in the Y-direction; (b) variation with antenna height in the Z-direction.

3.3.4 Effect of inter-turn spacing on absorption efficiency

Under the condition of a fixed current-strip length, this study investigates the influence of the inter-turn spacing of the strip in SA on its absorption efficiency. Figure 16 shows the variation of absorption efficiency with different inter-turn spacings when the strip lengths (L) are 1.0 m and 0.6 m, respectively. The simulation results indicate that the absorption efficiency exhibits a nonlinear relationship with the inter-turn spacing: when the inter-turn spacing is small, the absorption efficiency increases with increasing spacing; however, when the spacing exceeds a certain critical value, its influence gradually saturates. Preliminary analysis suggests that this critical spacing is closely related to the distance between the end of the current-strip and the sidewall of the antenna box. When the radial spacing is small, the end of the strip is far from the antenna box sidewall, resulting in a weak suppression effect from the box on electromagnetic radiation. As the spacing increases, the end of the strip gradually approaches the sidewall, and the suppression effect of the antenna box on the absorption efficiency becomes more pronounced, thereby offsetting the gain brought by the increased spacing. The critical spacing corresponding to different strip lengths in Figure 15 further validates the aforementioned mechanism: when the current-strip length is shorter (0.6 m), a larger inter-turn spacing is required for the strip end to achieve a similar absorbed state with the box wall as in the case of a longer strip. This result reveals the important influence of the relative configuration between the antenna and the box wall on the absorption characteristics.Furthermore, we speculate that the oscillation of the absorption efficiency in the saturation region may also be related to changes in the projection length of the current-strip parallel and perpendicular to the magnetic field caused by inter-turn spacing adjustments.

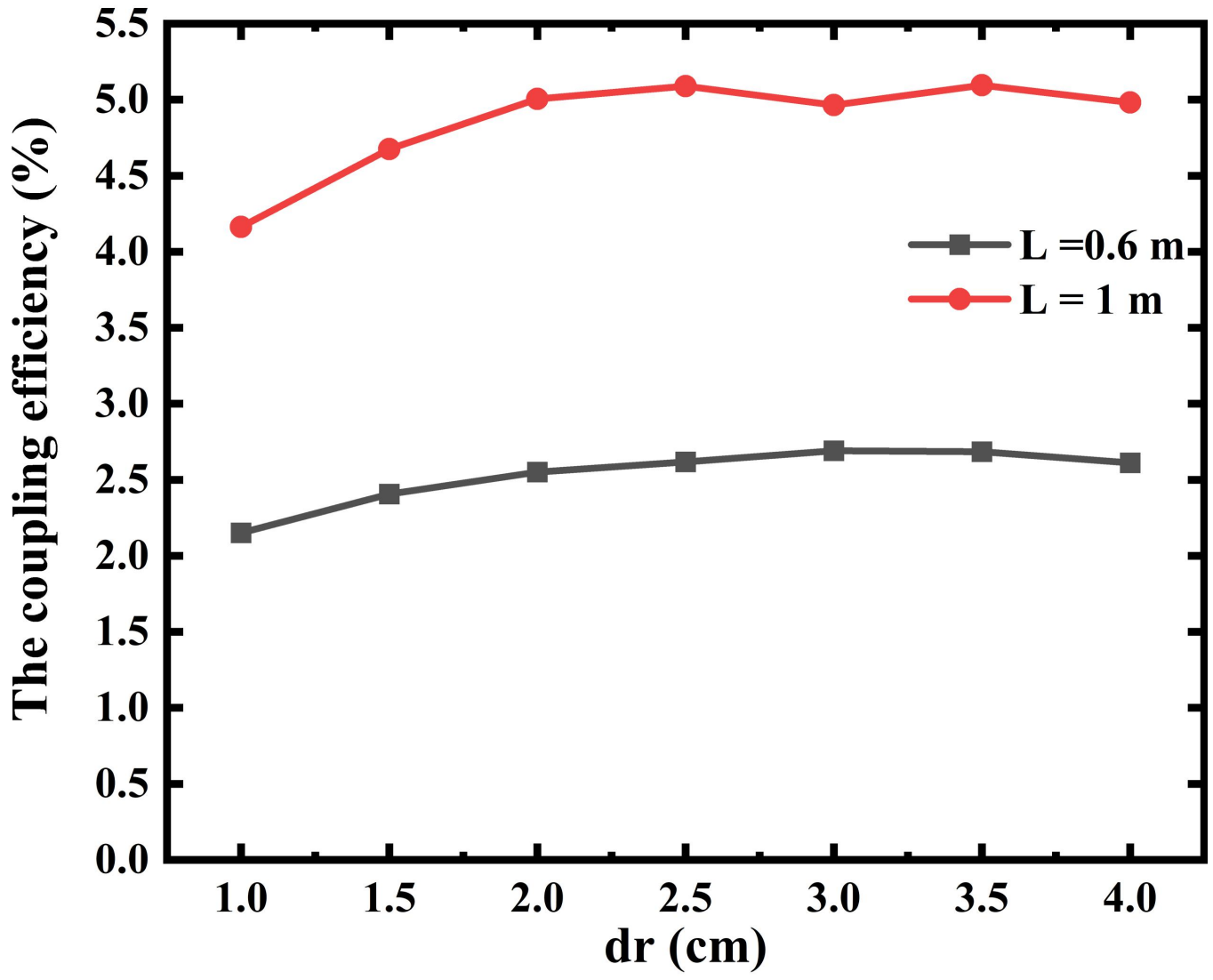

Figure 16. Effect of inter-turn spacing on the absorption efficiency of the SA.

3.3.5 Effect of current-strip width on absorption efficiency

Under the condition that both the length of the current-strip and the radial pitch remain unchanged, we investigated the influence of the coil radius of the spiral antenna on the antenna's absorption efficiency. With the radial pitch of the coil kept constant, increasing the width of the current-strip reduces the spacing between adjacent current-strips. However, Figure 17 still shows that the absorption efficiency of the antenna increases with the increase in the width of the current-strip. This not only indicates that a wider current-strip can enhance the absorption efficiency of the antenna, but also that the width has a better effect on improving the absorption efficiency than the spacing between current-strips. Due to geometric constraints, the current-strip width of the SA cannot exceed 2 cm.

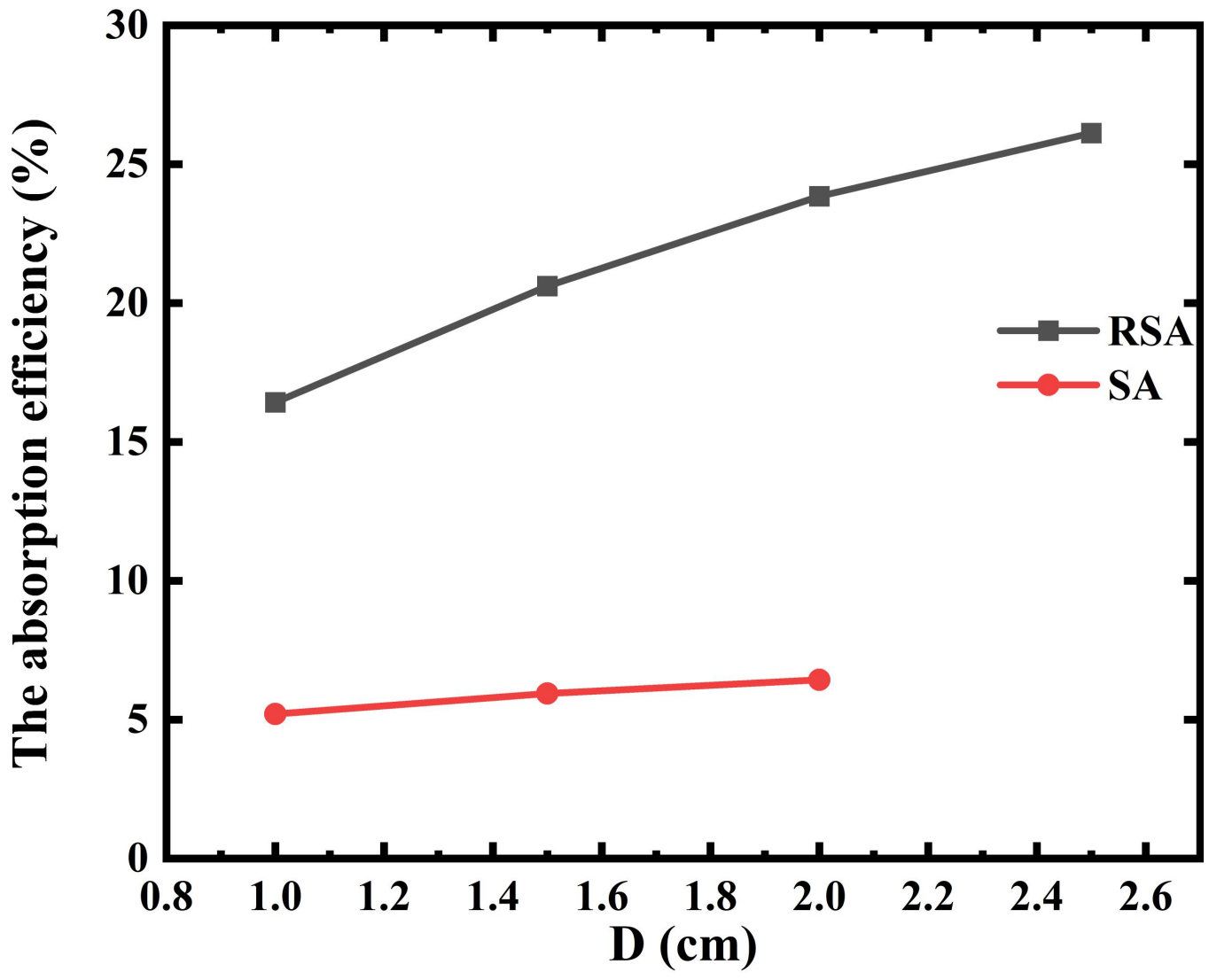

Figure 17. Effect of current-strip width on the absorption efficiency of the SA and RSA.

In summary, the effects of antenna terminal condition, current-strip length, inter-turn spacing, current-strip width, and antenna box size on the absorption efficiency in Helimak are as follows:

- Under low power conditions, the open-circuit design structure is relatively simple and generally exhibits higher absorption efficiency compared to a short-circuit design;
- The absorption efficiency increases approximately linearly with the length of the current-strip, reflecting the enlarged effective radiating area;
- There is a critical value in the design of the inter-turn spacing. When the inter-turn spacing is small, the absorption efficiency increases with the increase of the spacing, but it no longer increases significantly once the spacing exceeds a certain threshold;
- Increasing the current-strip width effectively enhances the absorption efficiency, and its optimization effect is superior to adjusting the inter-turn spacing;
- Enlarging the antenna box size (i.e., increasing the distance between the metal wall and the antenna) helps improve absorption efficiency, whereas excessively close wall proximity leads to a significant efficiency reduction.

### *3.4. Optimized antenna design*

According to the findings above, the design of antennas should prioritize parameter optimization in the following order: First, for low-power antennas used in boundary plasma absorption, the open-circuit design offers superior absorption efficiency. Second, a sufficient distance must be maintained between the antenna and the box wall to avoid near-field suppression effects. Third, extending the current-strip length should be prioritized, ideally covering the entire ceramic window area, as this is the most effective approach to enhancing power absorbed. Finally, when the length and coverage area are fixed, increasing the current-strip width contributes more to efficiency improvement than increasing the radial spacing. Therefore, provided that electric field breakdown or arc discharge does not occur, the strap width may be appropriately increased for further performance enhancement. Based on these design principles, we have developed an open-circuit racetrack-shaped spiral antenna for the new recessed window structure. The reference design parameters of the antenna include 4.75 turns and a current-strip width of 0.015 m. Figure 18 presents the simulated radial electric field distribution of the racetrack-shaped spiral antenna. The results show that the electric field distribution excited by this antenna in plasma regions is similar to that of the previously four antennas. The calculation results also show that the antenna's absorption efficiency reaches approximately 65%, representing an improvement of more than ten times compared to the conventional short-circuited RSA. These results effectively validate the rationality of the design methodology proposed in this study and provide a valuable reference for engineering design.

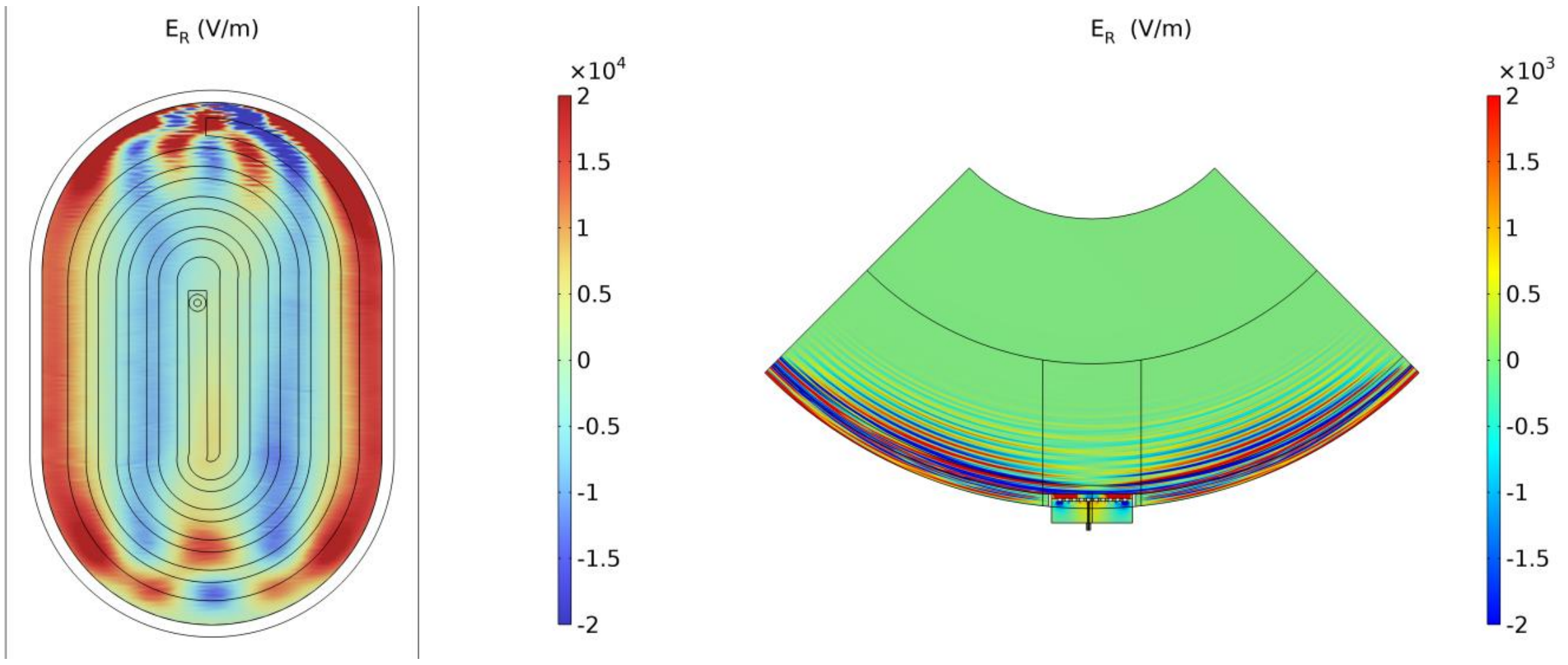

(a) (b)

Figure 18. Two-dimensional distributions of the electric field for the racetrack-shaped spiral antenna: (a) at the window–plasma interface; (b) at the antenna mid-plane.

## 4. Discussion

### *4.1 Fusion relevance and limitations of Helimak-based extrapolation*

Although the present study is motivated by applications to tokamak SOL density control and RF absorption optimisation, all simulations are performed using parameters characteristic of the Helimak device. Helimak operates in a low-temperature, low-density, electron-heated regime with a simple toroidal magnetic topology, which differs from reactor-relevant tokamak edge plasmas. It is therefore important to clarify the scope, relevance, and limitations of extrapolating the present results to tokamak conditions. Several key physical similarities justify the use of Helimak as a reduced model for exploring RF-plasma coupling mechanisms relevant to tokamak SOLs. These include the toroidal geometry, open magnetic field lines intersecting material boundaries, strong plasma gradients near the plasma–wall interface, and RF accessibility conditions typical of helicon and ion-cyclotron frequency ranges. As a consequence, Helimak provides a controlled platform to isolate fundamental coupling physics, such as the role of antenna geometry, near-field structure, and plasma density gradients, without the full complexity of a burning plasma environment. At the same time, important differences must be acknowledged. Compared with tokamak SOLs, Helimak plasmas exhibit much lower electron temperature, electron density and Magnetic field, higher effective collisionality, different neutral pressure and recycling conditions, negligible plasma β, and the absence of strong magnetic shear and X-point geometry. In addition, the present simulations neglect nonlinear plasma responses that may be significant in high-power tokamak RF operation. These differences preclude direct quantitative extrapolation of the absolute absorption efficiencies obtained here to tokamak devices. Accordingly, the results of this work should be interpreted as providing qualitative physical insight and design guidance, rather than predictive performance metrics for tokamaks. While the absolute values of absorption efficiency are specific to Helimak parameters, the observed trends—such as the sensitivity to antenna termination, current-strap length, inter-turn spacing, and near-field structure—are expected to remain relevant across a broader class of open-field-line plasmas. These trends highlight general principles that can inform the conceptual design and optimisation of RF antennas for tokamak SOL applications. Throughout this manuscript, statements referring to tokamak relevance are therefore intended in this restricted sense. Although this study does not establish a direct optimization strategy for tokamak operation, it reveals, through experimental and numerical analysis, the physical mechanisms of power deposition applicable to the tokamak edge and scrape-off layer (SOL) environment. These mechanisms must be interpreted with appropriate scaling considerations and due caution.

*4.2 Ordering and interpretation of RF damping mechanisms*

The relative importance of the different RF damping channels considered in this work—Landau damping, Doppler-shifted cyclotron damping, and anomalous Doppler damping—can be understood from the characteristic frequency and velocity orderings under Helimak conditions. For the parameter range relevant to the present simulations, the electron cyclotron frequency is typically $\omega_{ce}\sim 10^9{-}10^{10}\,\mathrm{s}^{-1}$, while the parallel thermal velocity satisfies $\upsilon_{Te}=\sqrt{2T_e/m_e}\sim (1{-}3)\times10^6\,\mathrm{m\,s}^{-1}$ for $T_e=5{-}15\,\mathrm{eV}$. With parallel wavenumbers $k_\parallel\sim 1{-}10\,\mathrm{m}^{-1}$, the characteristic Doppler frequency satisfies $k_\parallel\upsilon_{Te}\sim 10^6{-}10^7\,\mathrm{s}^{-1}$, yielding the ordering $\omega_{ce}\gg k_\parallel\upsilon_{Te}$. Under this condition, resonant cyclotron interactions are strongly suppressed, explaining the negligible contribution of Doppler-shifted cyclotron damping in the present results. Similarly, anomalous Doppler damping requires the simultaneous satisfaction of the resonance condition $\omega-k_\parallel\upsilon_{Te}+\omega_{ce}\approx0$ and sufficiently weak collisional decorrelation. In Helimak, the effective electron collision frequency, dominated by electron–neutral collisions, is typically $\nu_e\sim 10^5{-}10^6\,\mathrm{s}^{-1}$, comparable to $k_\parallel\upsilon_{Te}$. This collisionality further suppresses anomalous Doppler damping by broadening the resonance and reducing phase coherence. Consequently, anomalous Doppler damping contributes negligibly to the total power absorbed under the conditions studied here. The dominance of Landau damping in the present simulations does not arise simply from the low electron temperature, but rather from the strong parallel electric fields associated with the TG mode excited by the antenna. These modes possess short parallel wavelengths and substantial $E_\parallel$, which efficiently couple to electrons satisfying the Landau resonance condition $\omega\approx k_\parallel\upsilon_{Te}$. As a result, Landau damping remains the primary damping channel even in a weakly magnetised, low-temperature plasma. It should be emphasised that while antenna geometry strongly influences the near-field structure and therefore the absorption efficiency, it has only a secondary effect on the partition of damping mechanisms under the Helimak conditions considered here. For fixed plasma parameters and operating frequency, variations in antenna geometry primarily modify the amount of power coupled into TG modes, while the subsequent dissipation remains dominated by Landau damping. This conclusion is specific to the present parameter regime and does not preclude geometry-dependent changes in damping partition under different plasma conditions.

*4.3 Interpretation and limitations of antenna optimisation results*

Sections 3.3–3.4 present a systematic numerical optimisation of antenna geometry, including current-strap length, inter-turn spacing, conductor width, and the transition from spiral to racetrack-shaped designs. The following statements are based directly on numerical evidence obtained from the present simulations: (i) open-circuit termination consistently yields higher absorption efficiency than short-circuit termination; (ii) increasing current-strap length leads to a monotonic increase in coupled power over the range studied; and (iii) under identical plasma conditions, the racetrack-shaped spiral antenna achieved the highest absorption efficiency among all configurations considered, representing an improvement of approximately one order of magnitude compared to the reference RSA geometry. These findings are reproducible across the parameter scans presented and are therefore regarded as robust numerical results. Interpretation of the underlying physical mechanisms is necessarily more tentative. Given the magnitude of the reported improvement, potential numerical artefacts were carefully considered. In addition, sensitivity tests varying the size of the computational domain and the placement of absorbing boundaries showed no qualitative change in the relative ranking of antenna configurations. These checks indicate that the observed optimisation trends are not dominated by numerical discretisation or boundary effects within the parameter space explored. Nevertheless, the present optimisation study is subject to important limitations. The simulations are linear and electromagnetic in nature and do not address engineering constraints such as RF breakdown, arcing limits, conductor heating, mechanical tolerances, or long-pulse thermal loads. As a result, the optimised antenna geometries identified here should be interpreted as demonstrating electromagnetic absorbed potential, rather than as directly deployable designs. Incorporating these additional constraints, as well as nonlinear plasma responses at higher power, remains an important topic for future work.

## 5. Conclusion

In this work, we developed THEMIS to investigate helicon wave propagation, damping, and antenna–plasma coupling in the Helimak device. By incorporating realistic magnetic-field and density profiles and employing a finite-temperature thermal dielectric tensor, the model reproduces the key features of wave accessibility in toroidal edge plasmas.

Consistent with dispersion analysis, the simulations show that fast-wave propagation is strongly suppressed under Helimak conditions, whereas slow-wave (TG) modes can propagate and deposit power in localized regions. The dominant heating channels—electron Landau damping and collisional damping—account for more than 95% of total absorption, in agreement with expectations for low-temperature, magnetized edge plasmas. Comparison of the four helicon wave antenna geometries used in recent Helimak experiments reveals that, under the present protruding-window configuration, helicon waves remain confined to the waveguide region, with negligible penetration into the core-accessible plasma. This finding is consistent with direct probe measurements, which showed no measurable increase in main-chamber density during the initial helicon wave heating campaign. The simulations identify two primary limiting mechanisms: rapid slow-wave cutoff in the low-density SOL and geometric trapping produced by the protruding launch structure. Introducing a recessed-window configuration fundamentally alters these constraints by placing the dielectric window directly inside the vacuum chamber, thereby reducing the evanescent-layer width and improving direct access to the boundary plasma. Parametric scans of the window recess distance demonstrate a clear non-monotonic dependence of absorption efficiency on launch position, providing quantitative guidance for experimental installation. Further scans of helical-type antenna designs reveal the physics-based optimization principles governing slow-wave excitation in toroidal geometry: the advantage of open-circuit termination at low power, the importance of maximizing current-strap length and width, the existence of a critical inter-turn spacing, and the need to minimize near-field suppression by maintaining adequate clearance from metallic walls. Leveraging these principles, we designed an optimized racetrack-shaped spiral antenna that increases the absorption efficiency by more than an order of magnitude compared with the short-circuited rectangular spiral antenna used in the first Helimak campaign. This improvement directly supports forthcoming experimental upgrades, where the recessed-window launcher and optimized antenna will be installed and validated against interferometry, probe, and RF diagnostic measurements.

Overall, this study establishes the physical basis and practical feasibility of helicon-assisted SOL density control as a pathway to enhanced RF absorption in fusion devices. The modelling framework is directly extendable to predictive studies on tokamaks equipped with ICRH and ECRH systems—such as J-TEXT, EAST, and future ITER-class devices—and can be integrated into OMFIT (or IMAS) workflows for reactor-relevant scenario development. Results motivate coordinated experiment–simulation campaigns to test helicon-driven edge density modulation and to quantify its potential benefits for ICRH matching, impurity control, and edge stability in magnetic-confinement fusion plasmas.

**Acknowledgements**

Inspiring discussions with Xing-Quan Wu and Ran Chen are appreciated. This work was supported by the National Natural Science Foundation of China (Grants No. 92271113, 12411540222, 12481540165, 12405274), the Natural Science Foundation Project of Chongqing (Grant No. CSTB2025NSCQ-GPX0725), the ENN Hydrogen–Boron Fusion Research Fund (Grant No. 2025ENNHB01-011), the Natural Science Foundation of Hunan Province (Grant No.2025JJ50006), the Basic Research Fund of the National Key Laboratory of Particle Transport and Enrichment Technology (Grant No. WZKF-2025-04), the National Key R&D Program of China (Grant No. 2022YFE03070000, 2022YFE03070004), the Anhui Provincial Key Research and Development Project (Grant No. 2023a05020008), and the Strategic Priority Research Program of the Chinese Academy of Sciences (Grant No. XDB0790302).